\documentclass[aps,epsfig,nofootinbib,a4paper,12pt,floatfix]{revtex4}
\usepackage[utf8]{inputenc}
\usepackage{lmodern}
\usepackage{blindtext}
\usepackage[T1]{fontenc}
\usepackage[dvipsnames]{xcolor}
\usepackage[utf8]{inputenc}
\usepackage[english]{babel}
\usepackage[letterpaper]{geometry}
\usepackage{graphicx}
\usepackage{clipboard}
\newclipboard{myclipboard}
\usepackage{lipsum}
\usepackage[english]{babel}
\usepackage[english]{babel}

\addto\captionsenglish{%
}

\usepackage{hyperref}
\usepackage{setspace}
\usepackage{float}
\usepackage{slashed}
\usepackage{amsmath}
\usepackage{epstopdf}
\usepackage{latexsym}
\usepackage{amssymb}
\usepackage[center]{subfigure}
\DeclareMathOperator{\arctanh}{arctanh}
\begin{document}
\begin{flushleft}
{\small\bf BU-HEPP-19-05}
\end{flushleft}
\vspace{18mm}
\title{IR-improved DGLAP parton shower effects for associated production of a W boson and jets in pp collisions at $\sqrt{s}=$8 and 13 TeV}

\author{B. Shakerin$^{a}$}
\email{bahram.shakerin@gmail.com}

\author{B.F.L. Ward$^{a}$}
\email{bfl_ward@baylor.edu}

\affiliation{$^{a}$Physics Department, Baylor
University, Waco, TX 76798-7316, USA}
\date{\today}
\begin{abstract}
In a previous paper, hereafter referred to as I, we have analyzed the 7 TeV LHC data on W + jets events from the standpoint of IR-improved DGLAP parton shower effects, using the IR-improved Herwiri1.031 parton shower MC in comparison with the Herwig6.5 parton shower MC in the context of the exact $O(\alpha_s)$ matrix element matched parton shower framework provided by MG5\_aMC@NLO. In the current paper, we extend this analysis to the LHC 8 and 13 TeV data to investigate the energy dependence of the results obtained in I. Specifically, W~+ jet events are generated in the MADGRAPH5\_aMC@NLO framework and showered by HERWIG6.521 and HERWIRI1.031 with $\mathtt{PTRMS}=2.2$ and 0~GeV, respectively. The differential cross sections are reported as functions of jet multiplicity, transverse linear momenta ($P_{T}$), the jet pesudo-rapidity ($\eta$) and the scalar sum of jet transverse momenta ($H_{T}$) for different jet multiplicities 1--3. The dijet cross sections as functions of transverse linear momenta, invariant mass of the dijet and the jet separation are shown as well. Distributions of angular correlations between the jets and the muon are examined as well and the corresponding cross sections are presented. The respective measured cross sections are compared with the exact next-to-leading-order (NLO) matrix element matched parton shower theoretical predictions provided by MADGRAPH5\_aMC@NLO/HERWIRI1.031~($\mathtt{PTRMS}=0)$ and MADGRAPH5\_aMC@NLO/HERWIG6.521~($\mathtt{PTRMS}=2.2~\mathrm{GeV})$ and the phenomenological consequences are discussed with an eye toward their energy dependence. 
\end{abstract}
\maketitle
\thispagestyle{plain}
\pagestyle{plain}
\section{Introduction}
In this paper, we continue our analysis of the LHC W+ jets data which  we started in Ref.~\cite{bsh1} (hereafter referred to as I) with the data at the center of momentum system energy of 7 TeV. In I, we found that, for a large number of observables in the 
7 TeV data on $W+ n~ {\rm jets},~n=1,2,3$, the predictions from the exact NLO matrix element matched parton shower 
with IR-improvement via Herwiri1.031~\cite{Joseph:2010cq} in the MADGRAPH5\_aMC@NLO~\cite{Alwall:2014hca} framework are as good or better in the soft regime than those in the same framework with the unimproved shower of Herwig6.5~\cite{Corcella:2000bw} where the latter shower has an intrinsic Gaussian transverse momentum distribution with a root-mean-squared value of 2.2GeV/c for the partons in the proton. In what follows, we consider the comparison between the LHC 8 and 13 TeV W+ jets data~\cite{Khachatryan:2016fue,Sirunyan:2017wgx} and the analogous sets of IR-improved and unimproved predictions from MADGRAPH5\_aMC@NLO/Herwiri1.031~($\mathtt{PTRMS}=0)$ and MADGRAPH5\_aMC@NLO/HERWIG6.521~($\mathtt{PTRMS}=2.2~\mathrm{GeV})$, respectively, in an obvious notation.\par
 In particular, we look into the energy dependence of our results by comparing the results from all three energies: 7, 8 and 13 TeV. For each of observable, we investigate the interplay of IR-improvement in the soft regime with the colliding beam energy. In this way, we elucidate the energy dependence of IR-improvement in the processes under study here.

   The theoretical background for our discussion is given in I and Ref. \cite{Shakerin:2017hbd} so that we do not repeat that here. In Section II, we describe the event generation, analysis methods and cuts that we use in our discussion. In Section III we present the results of our comparisons with the LHC 8 and 13 TeV data for several observables. In Section IV, we study the energy dependence of results provided here and in I, and Section V contains our concluding remarks.\par

\section{Event generation, Analysis and Cuts}
The generators for W~+ jet events are MADGRAPH5\_aMC@NLO \cite{Alwall:2014hca} interfaced with HERWIG6.521 and MADGRAPH5\_aMC@NLO interfaced with HERWIRI1.031, which use exact QCD next-to-leading-order (NLO) matrix element calculations. The number of events generated for the W, W~+~1 jet, W~+~2 jets, and W~+~3 jets processes are $10^7$, $10^6$, $10^5$, and $10^5$, respectively.\,These events are showered by MADGRAPH5\_aMC@NLO/HERWIRI1.031 (PTRMS = 0) and MADGRAPH5\_aMC@NLO/HERWIG6.521 (PTRMS = 2.2 GeV).\footnote{We will see later that HERWIRI gives either a better fit to the data or an acceptable fit without this extra Gaussian kick.}  During the analysis, jets were reconstructed using the anti-$k_{t}$ algorithm with FastJet \cite{Cacciari:2011ma} and the cuts in Table~\ref{t1} and Table ~\ref{t2} were imposed.

\begin{table}[h!]
\centering 
\begin{tabular}{ p{6cm}p{6cm} }
\hline
\multicolumn{2}{c}{Muon channel~~~($W\rightarrow \mu+\nu_{\mu}$)} $\sqrt{s}=8$ TeV\\
\hline
Lepton $P^{\mu}_{T}$ & $P^{\mu}_{T}>25~\mathrm{GeV}$  \\

Lepton rapidity $\eta_{\mu}$ & $|\eta_{\mu}|<2.1$  \\

Missing transverse energy  &$E^{\mathrm{miss}}_{T}>25~\mathrm{GeV}$  \\

Transverse mass    &$m_{T}>50~\mathrm{GeV}$ \\

Jet algorithm & Anti-$k_{t}$\\

Radius parameter~$R$& $R=0.5$\\

Jet $P^{jet}_{T}$ & $P^{\mathrm{jet}}_{T}>30~\mathrm{GeV}$ \\

Jet rapidity $\eta_{\mathrm{jet}}$ & $|\eta_{\mathrm{jet}}|<2.4$  \\

Jet isolation& $\Delta R(\mu,\mathrm{jet})>0.5$~(jet is removed)\\
\hline
\end{tabular}
\caption{Kinematic criteria defining the fiducial phase space for the $W\rightarrow \mu+\nu_{\mu}$~~channel }
\label{t1}
\end{table}
\begin{table}[h!]
\centering 
\begin{tabular}{ p{6cm}p{6cm} }
\hline
\multicolumn{2}{c}{Muon channel~~~($W\rightarrow \mu+\nu_{\mu}$)} $\sqrt{s}=13$ TeV\\
\hline
Lepton $P^{\mu}_{T}$ & $P^{\mu}_{T}>25~\mathrm{GeV}$  \\

Lepton rapidity $\eta_{\mu}$ & $|\eta_{\mu}|<2.4$  \\
Missing transverse energy  &$E^{\mathrm{miss}}_{T}>25~\mathrm{GeV}$  \\

Transverse mass    &$m_{T}>50~\mathrm{GeV}$ \\

Jet algorithm & Anti-$k_{t}$\\

Radius parameter~$R$& $R=0.4$\\

Jet $P^{jet}_{T}$ & $P^{\mathrm{jet}}_{T}>30~\mathrm{GeV}$ \\

Jet rapidity $\eta_{\mathrm{jet}}$ & $|\eta_{\mathrm{jet}}|<2.4$  \\

\hline
\end{tabular}
\caption{Kinematic criteria defining the fiducial phase space for the $W\rightarrow \mu+\nu_{\mu}$~~channel }
\label{t2}
\end{table}
The transverse mass, $m_{T}$, is defined as $m_{T}=\sqrt{2P^{\mu}_{T}P^{\nu_{\mu}}_{T}(1-\cos\Delta\phi})$ where $\Delta\phi$ is the difference in the azimuthal angle between the direction of the muon momentum and the associated muon neutrino, $\nu_{\mu}$, which can be written as
\begin{equation}
\Delta\phi=\phi^{\mu}-\phi^{\nu_{\mu}}.
\end{equation}
Rapidity is defined as $\displaystyle\frac{1}{2}\ln\left[\frac{E+p_{z}}{E-p_{z}}\right]$, where $E$ denotes the energy of the particle and $p_{z}$ is the longitudinal component of the momentum. Finally, the jet isolation, $\Delta R$, which is a Lorentz invariant quantity, is defined as 
\begin{equation}
\Delta R(\mu,\mathrm{jet})=\sqrt{\Delta\phi^2(\mu,\mathrm{jet})+\Delta\eta^2(\mu,\mathrm{jet})},
\end{equation}
where
\begin{equation}
\left\{ \begin{array}{ll}
         \Delta\phi(\mu,\mathrm{jet})=\phi_{\mu}-\phi_{\mathrm{jet}},\\
         \Delta\eta(\mu,\mathrm{jet})=\eta_{\mu}-\eta_{\mathrm{jet}},\\
         \eta=-\ln\tan(\frac{\theta}{2}),
         \end{array} \right. 
\label{etadef}
\end{equation}
where $\theta$  is the angle between the respective three-momentum vector and the positive beam direction.
\section{Results}
The measured W($\rightarrow\mu+\nu_{\mu}$)~+~jets cross sections \cite{Khachatryan:2016fue,Sirunyan:2017wgx} are shown and compared to the predictions of MADGRAPH5\_aMC@NLO/HERWIRI1.031
(PTRMS = 0) and MADGRAPH5\_aMC@NLO/HERWIG6.521 (PTRMS = 2.2 GeV). The 8 TeV MC data sample allows us to determine the cross sections for jet multiplicities up to 3 and to study the fiducial cross sections as functions of most kinematic observables for up to three jets. Each distribution is combined separately by minimizing a $\chi^2$ function. The factors applied to the theory predictions are summarized in Appendix A and Appendix B. 
We have used the following notation throughout this paper:
\begin{itemize}
 \item herwiri~$\equiv$~MADGRAPH5\_aMC@NLO/HERWIRI1.031 (PTRMS~=~0);
 \item herwig~ $\equiv$~ MADGRAPH5\_aMC@NLO/HERWIG6.521 (PTRMS~=~2.2~GeV).
\end{itemize}
\subsection{Results for $\sqrt{s}=$8 TeV}
\subsubsection{Transverse Momentum Distributions $P_{T}$}

The differential cross sections in jet $P_{T}$ for inclusive jet multiplicities from 1 to 3 are shown and compared with predictions provided by HERWIRI and HERWIG.
\par
\begin{figure}[H]
\centering
\includegraphics[scale=0.4]{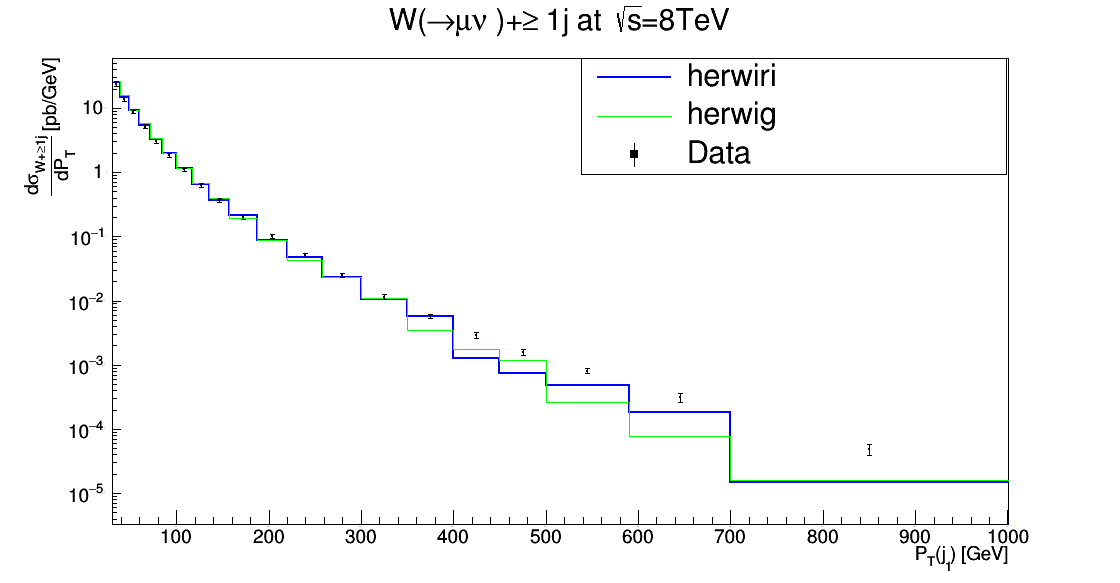}
\caption{Cross section for the production of W~+ jets as a function of the leading-jet $P_{T}$ in $N_{jet}\geq 1.$ The data are compared to predictions from MADGRAPH5\_aMC@NLO/HERWIRI1.031 and MADGRAPH5\_aMC@NLO/HERWIG6.521.}
\label{fig1}
\end{figure}
\begin{figure}[H]
\centering
\includegraphics[scale=0.4]{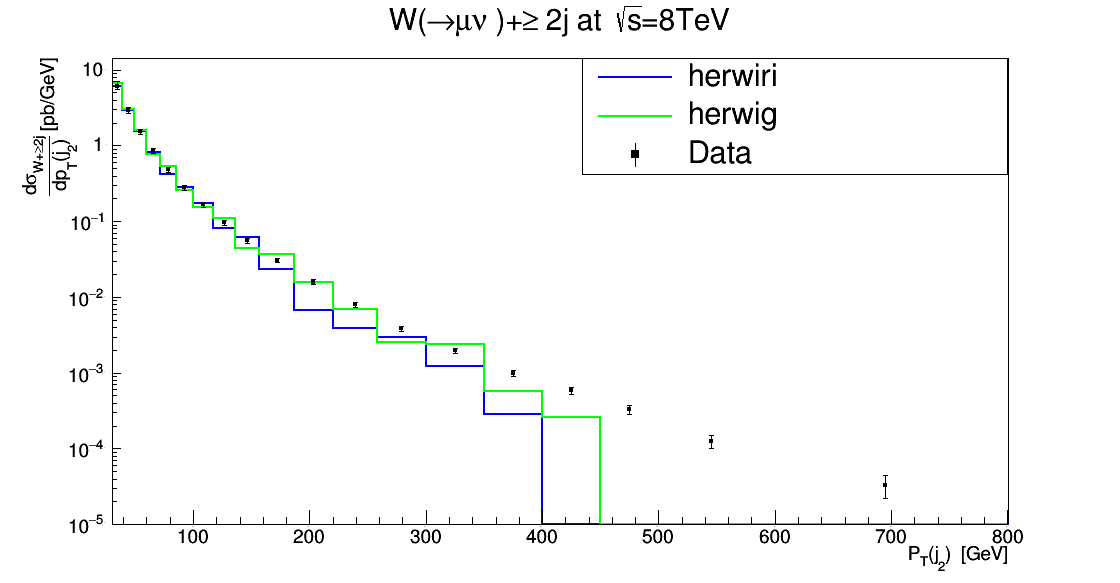}
\caption{Cross section for the production of W~+ jets as a function of the second leading-jet $P_{T}$ in $N_{jet}\geq2.$ The data are compared to predictions from MADGRAPH5\_aMC@NLO/HERWIRI1.031 and MADGRAPH5\_aMC@NLO/HERWIG6.521.}
\label{fig2}
\end{figure}
\begin{figure}[H]
\centering
\includegraphics[scale=0.4]{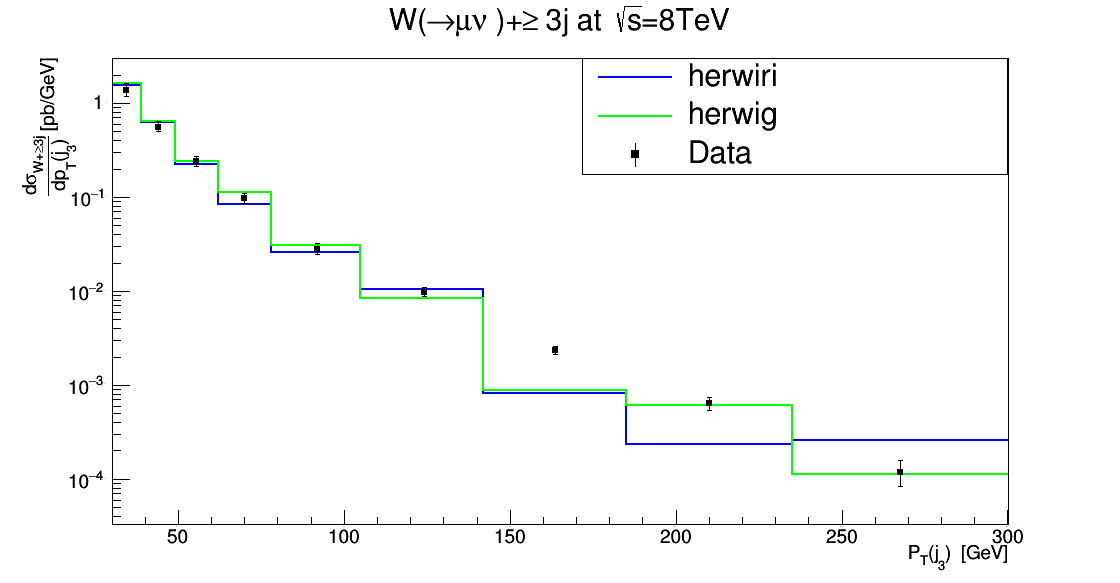}
\caption{Cross section for the production of W~+ jets as a function of the third leading-jet $P_{T}$ in $N_{jet}\geq 3.$ The data are compared to predictions from MADGRAPH5\_aMC@NLO/HERWIRI1.031 and MADGRAPH5\_aMC@NLO/HERWIG6.521.}
\label{fig3}
\end{figure}

The differential cross sections as functions of the first three leading jets are shown in Figure~\ref{fig1}, Figure~\ref{fig2}, and Figure~\ref{fig3}. In all three cases HERWIRI results give a better fit to the data in the soft regime. In Figure~\ref{fig1}, a good fit is provided by HERWIRI for $P_{T}<187~\mathrm{GeV}$ while for $P_{T}>400~\mathrm{GeV}$ the predictions provided by HERWIRI and HERWIG lie below the data. However, the HERWIRI predictions are closer to the data. In Figure~\ref{fig2} both HERWIRI and HERWIG provide a fairly good fit to the data for $P_{T}<100~\mathrm{GeV}$. In $100<P_{T}<300~\mathrm{GeV}$, there are cases in which the theoretical predictions provided by either HERWIRI or HERWIG overlap with the data. For higher values of $P_{T}$, $P_{T}>350~\mathrm{GeV}$, both HERWIRI and HERWIG underestimate the data although the HERWIG results are closer to the data in some cases.  In Figure~\ref{fig1}, for $P_{T}<187~\mathrm{GeV}$, $\big(\frac{\chi^2}{d.o.f}\big)_{\texttt{HERWIRI}}=0.59$ and $\big(\frac{\chi^2}{d.o.f}\big)_{\texttt{HERWIG}}=1.13$. In Figure~\ref{fig2}  for $P_{T}<140$~GeV, $\big(\frac{\chi^2}{d.o.f}\big)_{\texttt{HERWIRI}}=1.02$ and $\big(\frac{\chi^2}{d.o.f}\big)_{\texttt{HERWIG}}=1.95$. In Figure~\ref{fig3}, a very good fit is provided by HERWIRI to the data for $P_{T}<142~\mathrm{GeV}$. For higher values of $P_{T}$, HERWIG predictions overlap with the data while HERWIRI predictions either underestimates or overestimates the data. In Figure~\ref{fig3} $\big(\frac{\chi^2}{d.o.f}\big)_{\texttt{HERWIRI}}=1.47$ and $\big(\frac{\chi^2}{d.o.f}\big)_{\texttt{HERWIG}}=2.04$ for $P_{T}<142~\mathrm{GeV}$.\par
\vspace{1ex}

\noindent
\subsubsection{The Scalar Sum of Jet Transverse Momenta $H_{T}$ }
In this subsection, the differential cross sections are shown as function of $H_{T}$ for inclusive jet multiplicities 1--3. The scalar sum $H_{T}$ is defined as
\begin{equation}
H_{T}=\sum_{i=1}^{N_{\mathrm{jet}}}P_{T}(j_{i}),
\end{equation}
for each event.

\begin{figure}[H]
\centering
\includegraphics[scale=0.4]{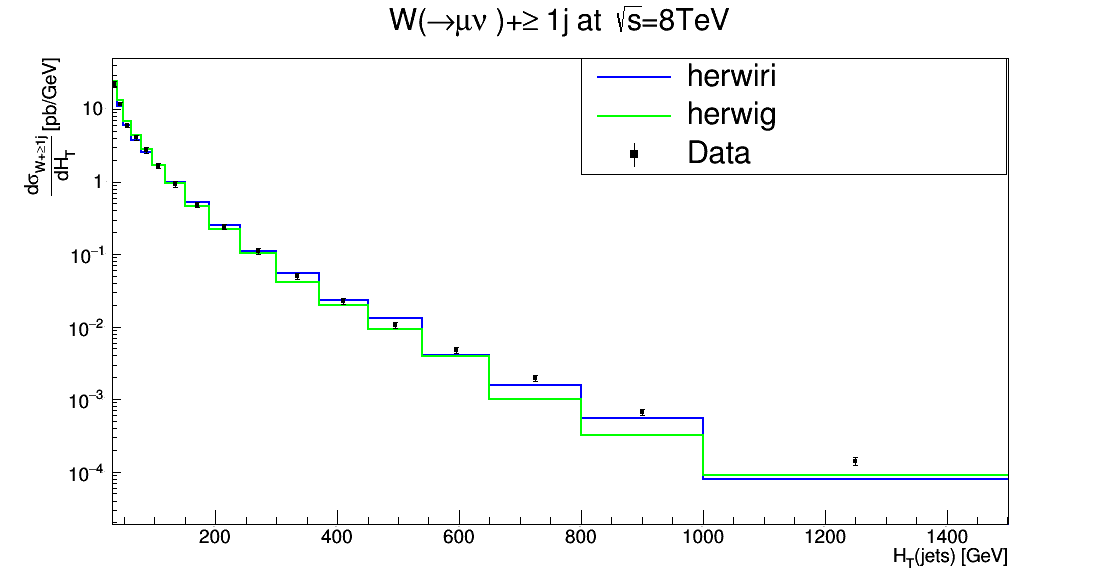}
\caption{Cross section for the production of W~+ jets as a function of  $H_{T}$ in $N_{jet}\geq 1.$ The data are compared to predictions from MADGRAPH5\_aMC@NLO/HERWIRI1.031 and MADGRAPH5\_aMC@NLO/HERWIG6.521.}
\label{fig4}
\end{figure}

\begin{figure}[H]
\centering
\includegraphics[scale=0.4]{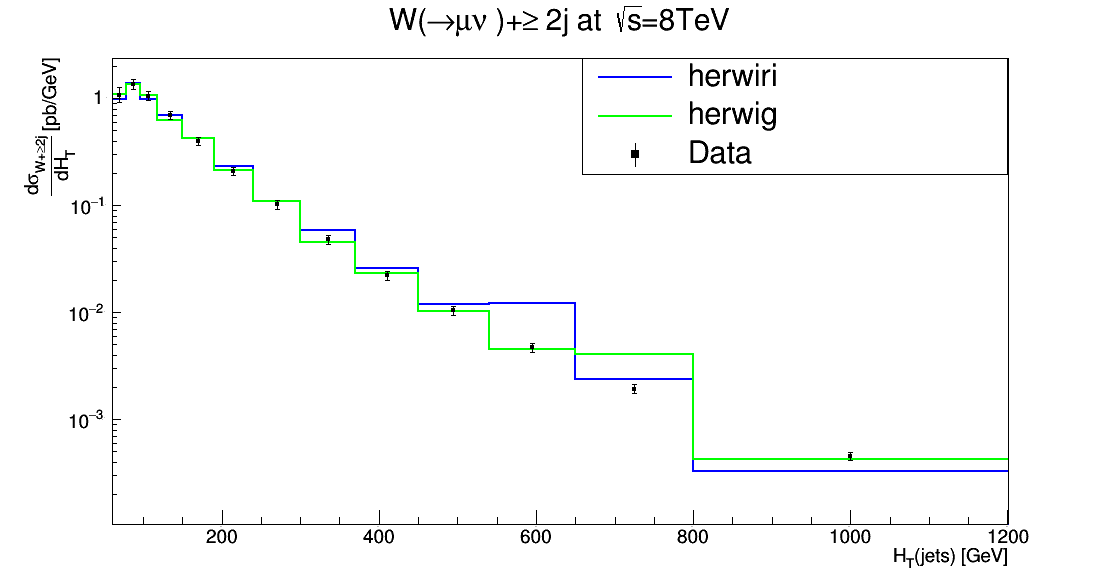}
\caption{Cross section for the production of W~+ jets as a function of  $H_{T}$ in $N_{jet}\geq2.$ The data are compared to predictions from MADGRAPH5\_aMC@NLO/HERWIRI1.031 and MADGRAPH5\_aMC@NLO/HERWIG6.521.}
\label{fig5}
\end{figure}
\begin{figure}[H]
\centering
\includegraphics[scale=0.4]{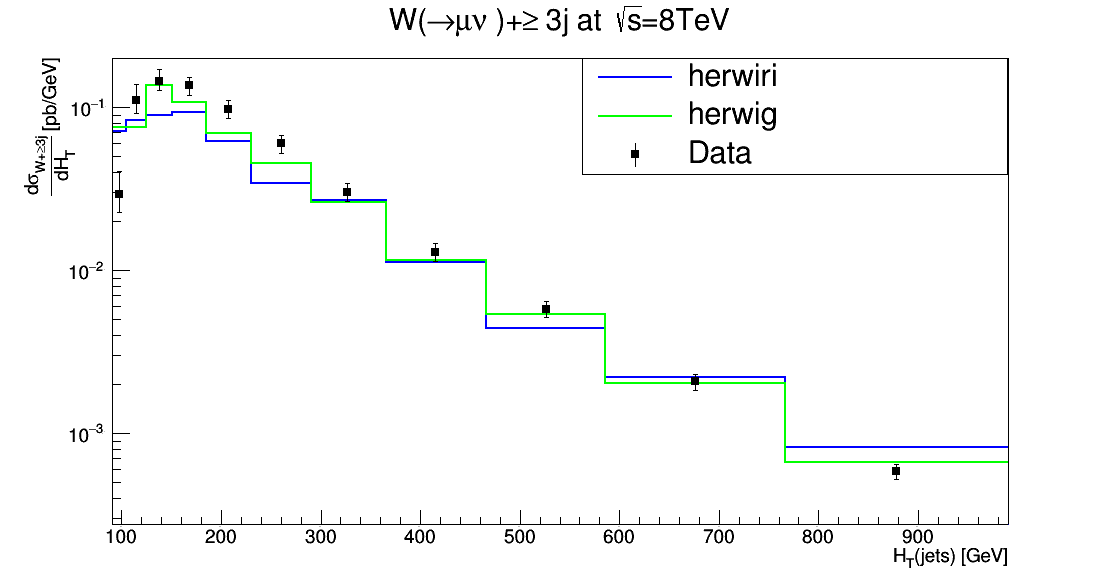}
\caption{Cross section for the production of W~+ jets as a function of $H_{T}$ in $N_{jet}\geq 3$. The data are compared to predictions from MADGRAPH5\_aMC@NLO/HERWIRI1.031 and MADGRAPH5\_aMC@NLO/HERWIG6.521.}
\label{fig6}
\end{figure}
The differential cross sections as functions of $H_{T}$ for inclusive jet multiplicities 1--3 are shown in Figure~\ref{fig4}, Figure~\ref{fig5}, and Figure~\ref{fig6}. In Figure~\ref{fig4}, a good fit is provided by HERWIRI predictions for $H_{T}<190~\mathrm{GeV}$. For higher values of $H_{T}$, HERWIRI predictions are closer to the data. In Figure~\ref{fig5}, in $H_{T}<190~\mathrm{GeV}$, HERWIRI gives a better fit to the data. For $190<H_{T}<600~\mathrm{GeV}$, HERWIG predictions overlap with the data while HERWIRI predictions overestimate the data. For higher values of $H_{T}$, in one case HERWIG overlaps and in one case HERWIRI overlaps with the data. In Figure~\ref{fig4}, for $H_{T}<190~~\mathrm{GeV}$, $\big(\frac{\chi^2}{d.o.f}\big)_{\texttt{HERWIRI}}=0.56$ and $\big(\frac{\chi^2}{d.o.f}\big)_{\texttt{HERWIG}}=1.58$. In Figure~\ref{fig5}, for $H_{T}<190~~\mathrm{GeV}$, $\big(\frac{\chi^2}{d.o.f})_{\texttt{HERWIRI}}=0.53$ and $\big(\frac{\chi^2}{d.o.f}\big)_{\texttt{HERWIG}}=0.82$. In Figure~\ref{fig6}, for $H_{T}<200~~\mathrm{GeV}$, $\big(\frac{\chi^2}{d.o.f}\big)_{\texttt{HERWIRI}}=13.20$ and $\big(\frac{\chi^2}{d.o.f}\big)_{\texttt{HERWIG}}=10.43$. Neither prediction is a good fit to the data in the soft regime in Fig.~\ref{fig6}.\par

\subsubsection{The Pseudorapidity Distributions $|\eta(j)|$}
In this section, the differential cross sections are shown as functions of pseudorapidities of the three leading jets. We note that pseudorapidity, which was defined in Eq.~(\ref{etadef}), can also be written as

\begin{equation}
\eta=\frac{1}{2}\ln(\frac{|\vec{P}|+P_{L}}{|\vec{P}|-P_{L}})=\arctanh(\frac{P_{L}}{|\vec{P}|}),
\end{equation}
where $\vec{P}$ is the particle three-momentum and $P_{L}$ is the component of the momentum along the beam axis. \par
\begin{figure}[H]
\centering
\includegraphics[scale=0.4]{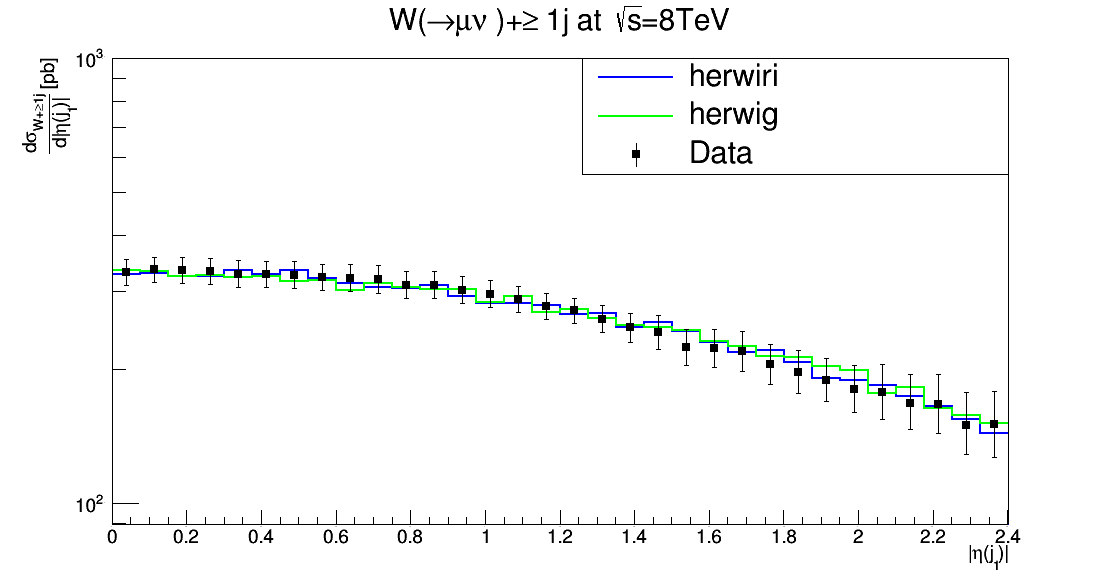}
\caption{Cross section for the production of W~+ jets as a function of  $|\eta(j_{1})|$ in $N_{jet}\geq 1.$ The data are compared to predictions from MADGRAPH5\_aMC@NLO/HERWIRI1.031 and MADGRAPH5\_aMC@NLO/HERWIG6.521.}
\label{fig7}
\end{figure}
\begin{figure}[H]
\centering
\includegraphics[scale=0.4]{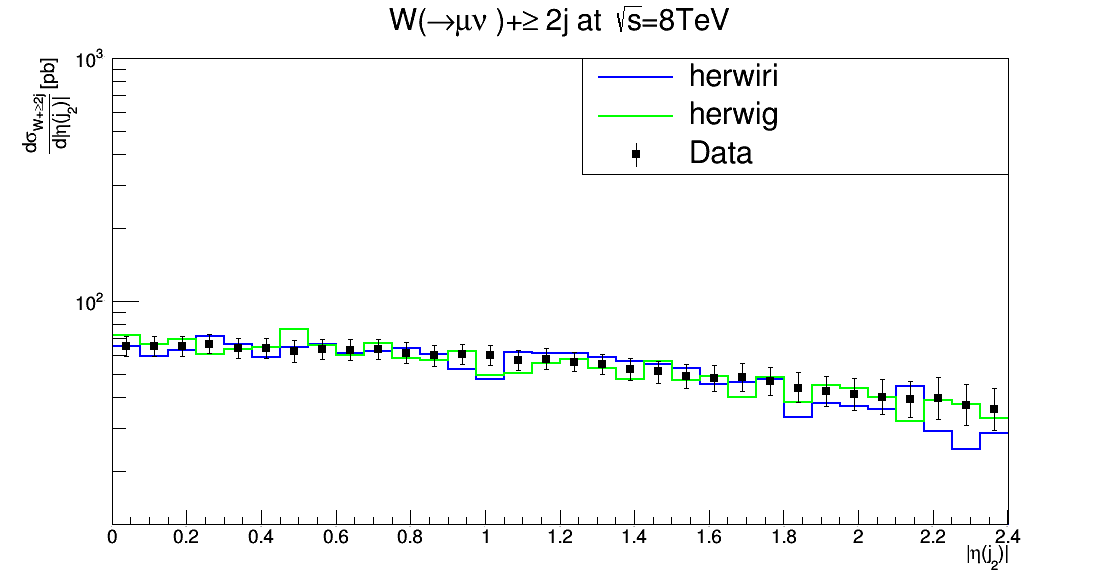}
\caption{Cross section for the production of W~+ jets as a function of  $|\eta(j_{2})|$ in $N_{jet}\geq2.$ The data are compared to predictions from MADGRAPH5\_aMC@NLO/HERWIRI1.031 and MADGRAPH5\_aMC@NLO/HERWIG6.521.}
\label{fig8}
\end{figure}

In Figure~\ref{fig7}, the cross sections are shown as a function of $|\eta(j_{1})|$, the leading jet pseudorapidity. The predictions provided by both HERWIRI and HERWIG give a very good fit to the data. In Figure~\ref{fig8}, the cross sections are shown as a function of $|\eta(j_{2})|$, the second leading jet pseudorapidity. The distribution is well modeled by both HERWIRI and HERWIG in $|\eta(j_{2})|<2.2$. For larger values of $|\eta(j_{2})|$, HERWIG clearly gives a better fit to the data. In Figure~\ref{fig7},  $\big(\frac{\chi^2}{d.o.f}\big)_{\texttt{HERWIRI}}=0.30$ and $\big(\frac{\chi^2}{d.o.f}\big)_{\texttt{HERWIG}}=0.38$. In Figure~\ref{fig8},  $\big(\frac{\chi^2}{d.o.f}\big)_{\texttt{HERWIRI}}=0.84$ and $\big(\frac{\chi^2}{d.o.f}\big)_{\texttt{HERWIG}}=0.66$.\par
\begin{figure}[H]
\centering
\includegraphics[scale=0.4]{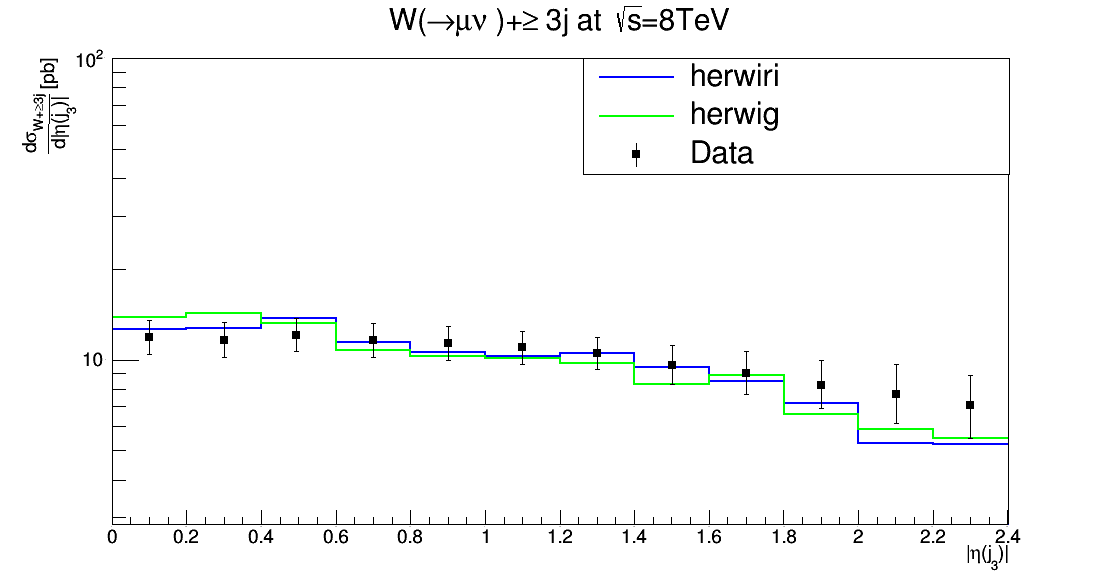}
\caption{Cross section for the production of W~+ jets as a function of $|\eta(j_{3})|$ in $N_{jet}\geq 3.$ The data are compared to predictions from MADGRAPH5\_aMC@NLO/HERWIRI1.031 and MADGRAPH5\_aMC@NLO/HERWIG6.521.}
\label{fig9}
\end{figure}

In Figure~\ref{fig9}, the cross sections are shown as a function of $|\eta(j_{3})|$, the third leading jet pseudorapidity. A very good fit is provided by HERWIRI for $|\eta(j_{3})|<2$. For higher values of $|\eta(j_{3})|$, both HERWIRI and HERWIG underestimate the data. However, the data are closer to HERWIG's predictions. In Figure~\ref{fig9},  $\big(\frac{\chi^2}{d.o.f}\big)_{\texttt{HERWIRI}}=0.62$ and $\big(\frac{\chi^2}{d.o.f}\big)_{\texttt{HERWIG}}=1.02$. \par

\subsubsection{Dijet $P_{T}$ Distributions}
In this section, the differential cross sections are shown as functions of the dijet $P_{T}$ (calculated from the two leading jets) for inclusive jet multiplicities 2--3. The dijet $P_{T}$ is defined as 
\begin{equation}
\mathrm{dijet}~ P_{T}=\sqrt{(P_{x}(j_{1})+P_{x}(j_{2}))^2+((P_{y}(j_{1})+P_{y}(j_{2}))^2},
\end{equation}
with
\begin{equation}
\left\{ \begin{array}{ll}
         j^{\mu}_{1}=(E_{j_{1}},P_{x}(j_{1}),P_{y}(j_{1}),P_{L}(j_{1})),\\
       j^{\mu}_{2}=(E_{j_{2}},P_{x}(j_{2}),P_{y}(j_{2}),P_{L}(j_{2})),\\\end{array} \right.
\end{equation}
where
\begin{equation}
P_{T}=\sqrt{P^2_{x}+P^2_{y}}.
\end{equation}
\begin{figure}[H]
\centering
\includegraphics[scale=0.4]{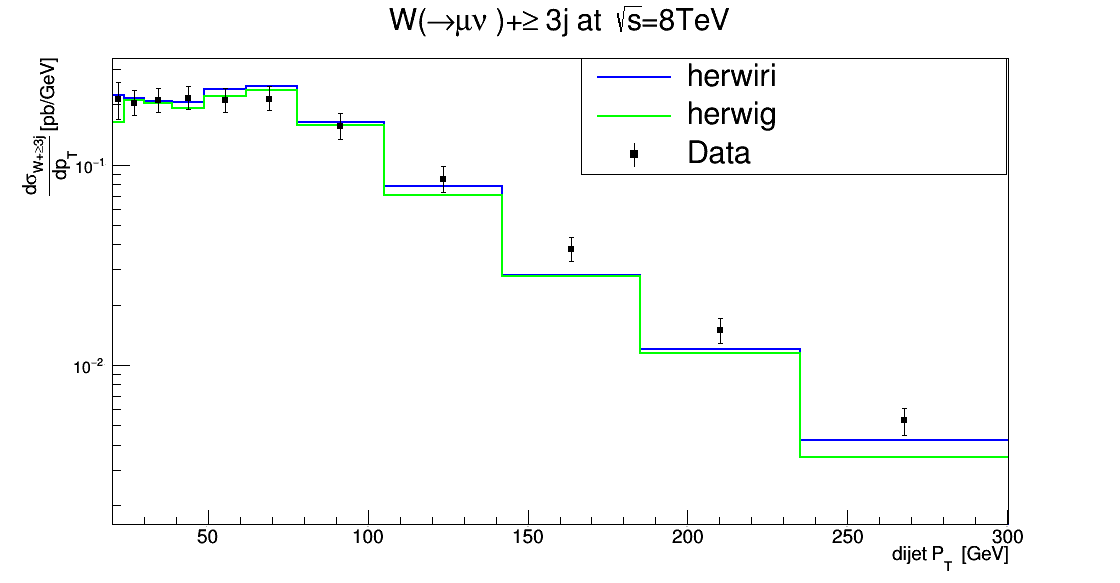}
\caption{Cross section for the production of W~+ jets as a function of  dijet $P_{T}$ in $N_{jet}\geq 3.$ The data are compared to predictions from MADGRAPH5\_aMC@NLO/HERWIRI1.031 and MADGRAPH5\_aMC@NLO/HERWIG6.521.}
\label{fig10}
\end{figure}

\begin{figure}[H]
\centering
\includegraphics[scale=0.4]{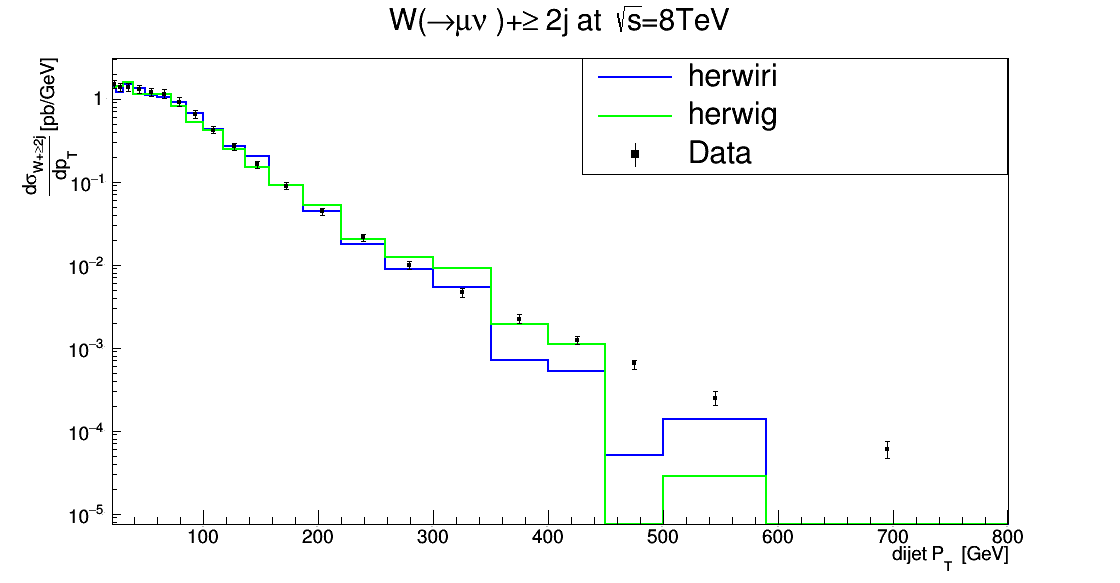}
\caption{Cross section for the production of W~+ jets as a function of  dijet $P_{T}$ in $N_{jet}\geq2.$ The data are compared to predictions from MADGRAPH5\_aMC@NLO/HERWIRI1.031 and MADGRAPH5\_aMC@NLO/HERWIG6.521.}
\label{fig11}
\end{figure}
In Figure~\ref{fig10}, the cross section is shown as function of the dijet $P_{T}$ for $N_{jet}\geq3.$ A better fit is provided for the data by the HERWIRI predictions in $P_{T}<150~\mathrm{GeV}$. For higher values of $P_{T}$, the predictions provided by both HERWIRI and HERWIG lie below the data points although HERWIRI is closer to the data. In Figure~\ref{fig10}, $\big(\frac{\chi^2}{d.o.f}\big)_{\texttt{HERWIRI}}=1.17$ and $\big(\frac{\chi^2}{d.o.f}\big)_{\texttt{HERWIG}}=1.43$.
In Figure~\ref{fig11}, the cross section is shown as function of the dijet $P_{T}$ for $N_{jet}\geq2.$ In this case again a better fit is provided by HERWIRI in $P_{T}<350~\mathrm{GeV}$. For $350<P_{T}<450~\mathrm{GeV}$, HERWIG gives a better fit to the data. For $P_{T}\geq450~\mathrm{GeV}$, the predictions provided by both HERWIRI and HERWIG underestimate the data, although HERWIRI results are closer to the data. In Figure~\ref{fig11}, $\big(\frac{\chi^2}{d.o.f}\big)_{\texttt{HERWIRI}}=2.07$ and $\big(\frac{\chi^2}{d.o.f}\big)_{\texttt{HERWIG}}=2.52$ for $P_{T}<187~\mathrm{GeV}$.
\subsubsection{The Rapidity Difference Distributions}
In this subsection, differential cross sections are presented as functions of the difference in rapidity. The difference in rapidity between the first and second leading jets is defined as 
\begin{equation}
|\Delta Y(j_{1},j_{2})=|Y(j_{1})-Y(j_{2})|,
\end{equation}
where
\begin{equation}
\left\{ \begin{array}{ll}
         Y(j_{1})=\displaystyle\frac{1}{2}\ln\left[\frac{E_{j_{1}}+P_{L}(j_{1})}{E_{j_{1}}-P_{L}(j_{1})}\right],\\[2ex]
        Y(j_{2})=\displaystyle\frac{1}{2}\ln\left[\frac{E_{j_{2}}+P_{L}(j_{2})}{E_{j_{2}}-P_{L}(j_{2})}\right],\end{array} \right.
\end{equation}
where $E_{j_{1}}$ and $E_{j_{1}}$ are energies for the first and the second leading jet, respectively. $P_{L}(j_{1})$ and $P_{L}(j_{2})$ represent the longitudinal momenta for the first and second leading jet.
\par
\begin{figure}[H]
\centering
\includegraphics[scale=0.4]{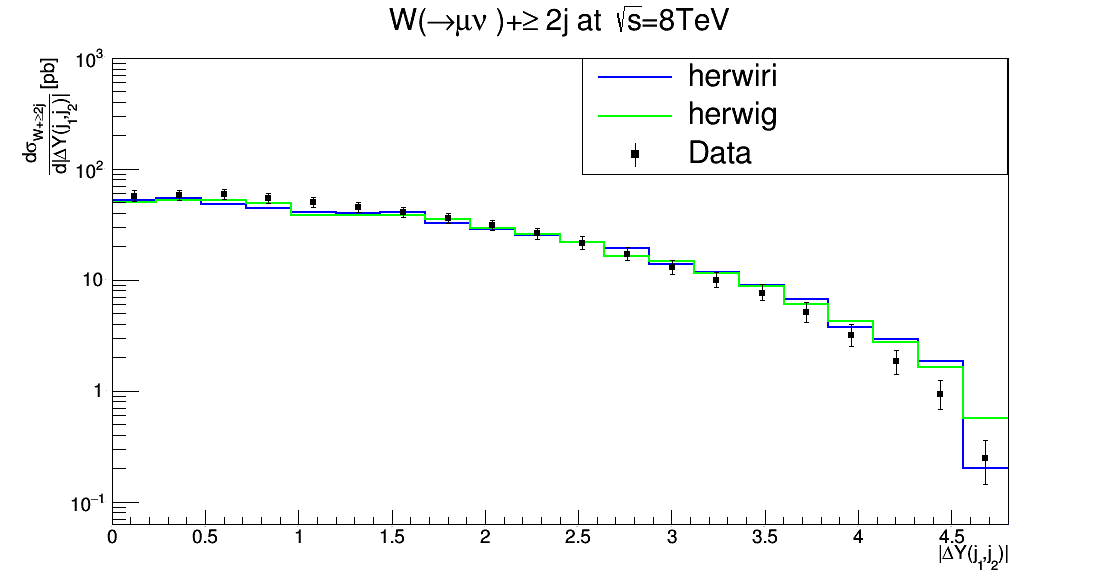}
\caption{Cross section for the production of W~+ jets as a function of difference is rapidity $|\Delta Y(j_{1},j_{2})|$ in $N_{jet}\geq 2.$ The data are compared to predictions from MADGRAPH5\_aMC@NLO/HERWIRI1.031 and MADGRAPH5\_aMC@NLO/HERWIG6.521.}
\label{fig12}
\end{figure}
\begin{figure}[H]
\centering
\includegraphics[scale=0.4]{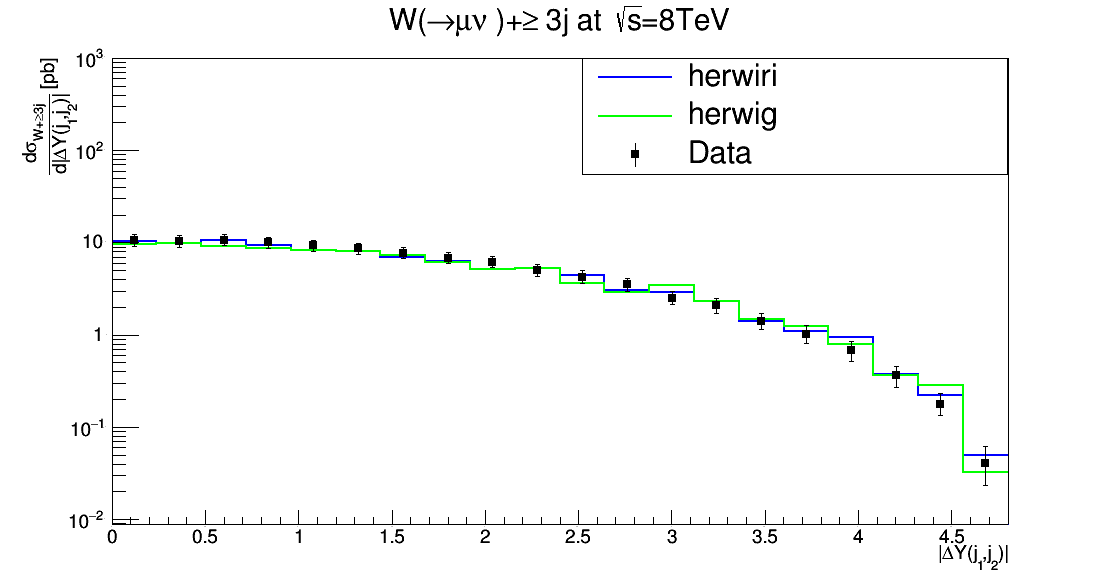}
\caption{Cross section for the production of W~+ jets as a function of  difference is rapidity $|\Delta Y(j_{1},j_{2})|$ in $N_{jet}\geq3.$ The data are compared to predictions from MADGRAPH5\_aMC@NLO/HERWIRI1.031 and MADGRAPH5\_aMC@NLO/HERWIG6.521.}
\label{fig13}
\end{figure}

In Figure~\ref{fig12} and Figure~\ref{fig13}, cross sections are presented as functions of difference in rapidity for inclusive jet multiplicities 2--3.
Figure~\ref{fig12} shows that for cases $|\Delta Y(j_{1},j_{2})|\leq 0.5$ and $1<|\Delta Y(j_{1},j_{2})|\leq 3.5$, both HERWIRI and HERWIG give good fits to the data. In $0.5<|\Delta Y(j_{1},j_{2})|<1$, a better fit is given to the data by the predictions provided by HERWIG. In Figure~\ref{fig12}, $\big(\frac{\chi^2}{d.o.f}\big)_{\texttt{HERWIRI}}=2.00$ and $\big(\frac{\chi^2}{d.o.f}\big)_{\texttt{HERWIG}}=1.98$.

In Figure~\ref{fig13}, the data is well modeled by the predictions provided by both HERWIRI and HERWIG although the theoretical predictions provided by HERWIRI are closer to the data in many cases. In Figure~\ref{fig13}, $\big(\frac{\chi^2}{d.o.f}\big)_{\texttt{HERWIRI}}=0.48$ and $\big(\frac{\chi^2}{d.o.f}\big)_{\texttt{HERWIG}}=1.04$.
\par
\begin{figure}[H]
\centering
\includegraphics[scale=0.4]{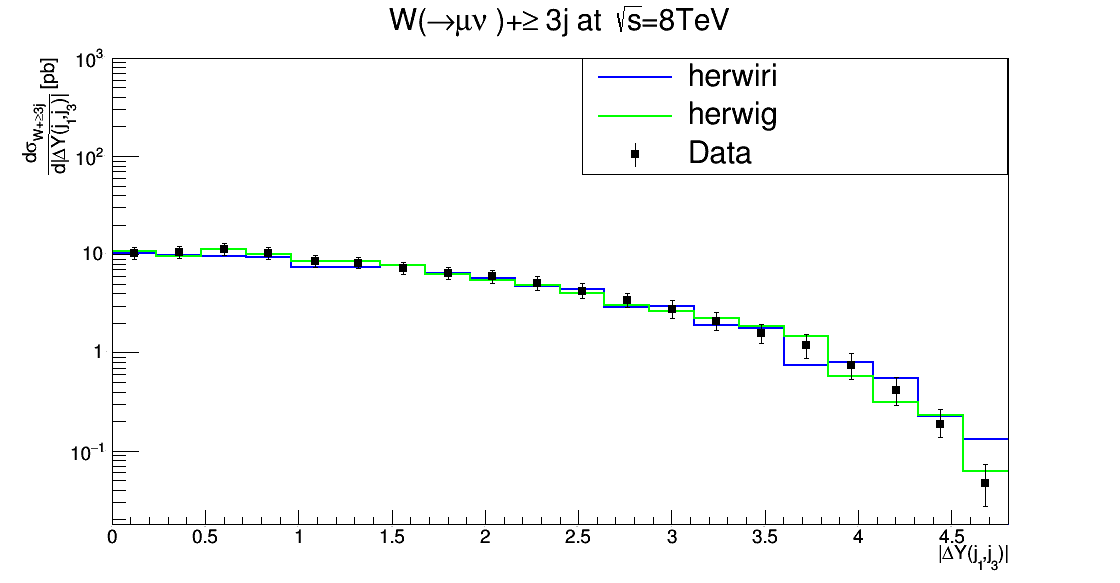}
\caption{Cross section for the production of W~+ jets as a function of difference is rapidity $|\Delta Y(j_{1},j_{2})|$ in $N_{jet}\geq 2.$ The data are compared to predictions from MADGRAPH5\_aMC@NLO/HERWIRI1.031 and MADGRAPH5\_aMC@NLO/HERWIG6.521.}
\label{fig14}
\end{figure}
\begin{figure}[H]
\centering
\includegraphics[scale=0.4]{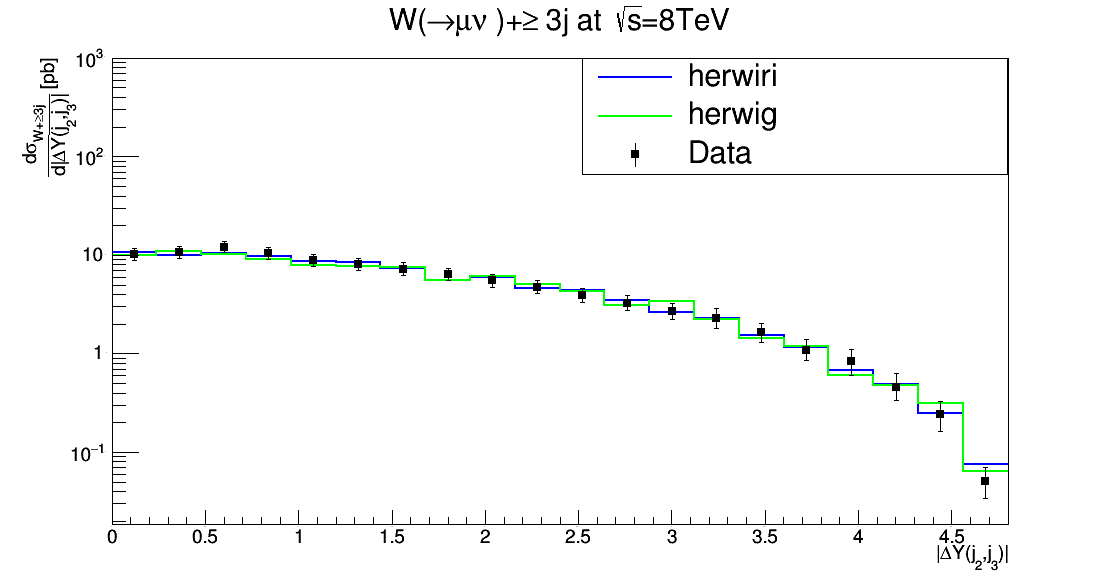}
\caption{Cross section for the production of W~+ jets as a function of  difference is rapidity $|\Delta Y(j_{2},j_{3})|$ in $N_{jet}\geq3.$ The data are compared to predictions from MADGRAPH5\_aMC@NLO/HERWIRI1.031 and MADGRAPH5\_aMC@NLO/HERWIG6.521.}
\label{fig15}
\end{figure}

In Figure~\ref{fig14} and Figure~\ref{fig15}, cross sections are presented as functions of difference in rapidity for inclusive jet multiplicity 3. In both cases, the data is well modeled by the predictions provided by both HERWIRI and HERWIG. In many cases HERWIRI predictions are closer to the data. In Figure~\ref{fig14}, $\big(\frac{\chi^2}{d.o.f}\big)_{\texttt{HERWIRI}}=1.20$ and $\big(\frac{\chi^2}{d.o.f}\big)_{\texttt{HERWIG}}=0.56$. In Figure~\ref{fig15}, $\big(\frac{\chi^2}{d.o.f}\big)_{\texttt{HERWIRI}}=0.33$ and $\big(\frac{\chi^2}{d.o.f}\big)_{\texttt{HERWIG}}=0.52$.
\vspace{10mm}

\subsubsection{Dijet Invariant Mass Distributions}
The cross sections are studied as functions of the dijet invariant mass calculated from the two leading jets for inclusive jet multiplicities 2--3. The dijet invariant mass is defined as
\begin{equation}
M({j_{1},j_{2}})=\sqrt{(E_{j_{1}}+E_{j_{2}})^2-(\vec{P}_{j_{1}}+\vec{P}_{j_{2}})^2}=\sqrt{m^2_{j_{1}}+m^2_{j_{2}}+2(E_{j_{1}}E_{j_{2}}-\vec{P}_{j_{1}}\cdot\vec{P}_{j_{2}})},
\end{equation}
where the leading jet is defined as $j^{\mu}_{1}=(E_{j_{1}}, \vec{P}_{j_{1}})$.
\par
\begin{figure}[H]
\centering
\includegraphics[scale=0.4]{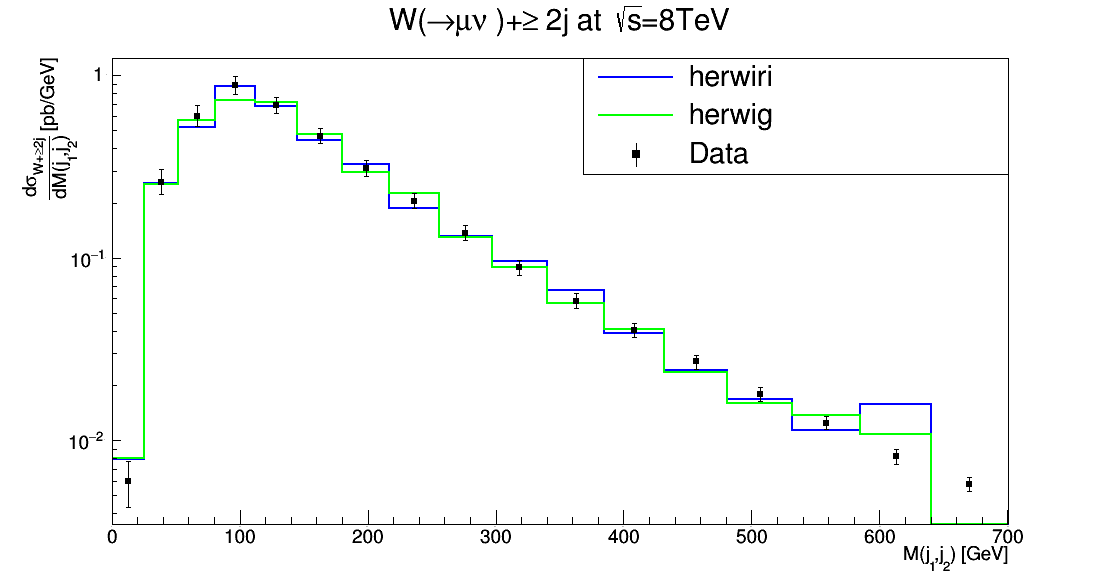}
\caption{Cross section for the production of W~+ jets as a function of dijet invariant mass $|M(j_{1},j_{2})|$ in $N_{jet}\geq 2.$ The data are compared to predictions from MADGRAPH5\_aMC@NLO/HERWIRI1.031 and MADGRAPH5\_aMC@NLO/HERWIG6.521.}
\label{fig16}
\end{figure}
\begin{figure}[H]
\centering
\includegraphics[scale=0.4]{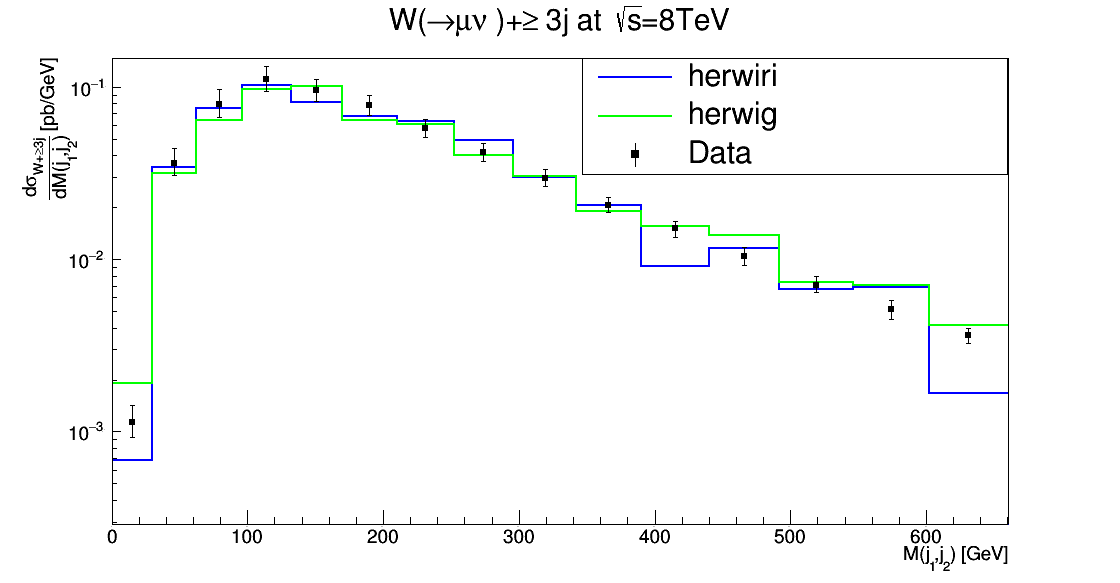}
\caption{Cross section for the production of W~+ jets as a function of  dijet invariant mass $|M(j_{1},j_{2})|$ in $N_{jet}\geq3.$ The data are compared to predictions from MADGRAPH5\_aMC@NLO/HERWIRI1.031 and MADGRAPH5\_aMC@NLO/HERWIG6.521.}
\label{fig17}
\end{figure}

In Figure~\ref{fig16} and Figure~\ref{fig17}, the cross sections are shown as functions of the dijet invariant mass for inclusive jet multiplicities 2--3. In Figure~\ref{fig16}, a good fit is provided by HERWIRI predictions to the data for $ M(j_{1},j_{2})<180~\mathrm{GeV}$ while for $200<M(j_{1},j_{2})<300~\mathrm{GeV}$, HERWIG gives a better fit to the data. For higher values of $ M$, the predictions provided by HERWIG are in better agreement with the data.

In Figure~\ref{fig17}, a better fit is provided by HERWIRI to the data for $50<M(j_{1},j_{2})<400~\mathrm{GeV}$. For higher values of $M(j_{1},j_{2})$, HERWIG predictions either overlap with the data or are closer to the data. In Figure~\ref{fig16}, for $M(j_{1},j_{2})<180~\mathrm{GeV}$, $\big(\frac{\chi^2}{d.o.f}\big)_{\texttt{HERWIRI}}=0.92$ and $\big(\frac{\chi^2}{d.o.f}\big)_{\texttt{HERWIG}}=1.01$. In Figure~\ref{fig17}, for $M(j_{1},j_{2})<200~\mathrm{GeV}$, $\big(\frac{\chi^2}{d.o.f}\big)_{\texttt{HERWIRI}}=1.10$ and $\big(\frac{\chi^2}{d.o.f}\big)_{\texttt{HERWIG}}=2.42$. 

\begin{figure}[H]
\centering
\includegraphics[scale=0.4]{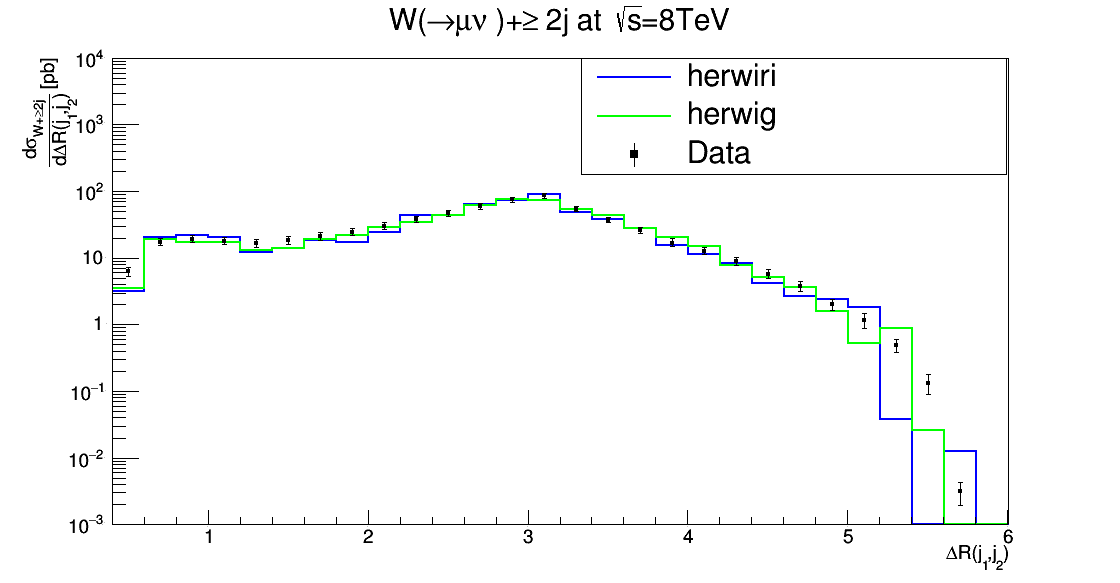}
\caption{Cross section for the production of W~+ jets as a function of the angular separation between the two leading jets $\Delta R(j_{1},j_{2})$ in $N_{jet}\geq 2.$ The data are compared to predictions from MADGRAPH5\_aMC@NLO/HERWIRI1.031 and MADGRAPH5\_aMC@NLO/HERWIG6.521.}
\label{fig18}
\end{figure}

Figure~\ref{fig18} shows the cross section as a function of angular separation between the first two leading jets for inclusive jet multiplicity 2. For $\Delta R({j_{1},j_{2}})<4.2$, HERWIRI predictions are in better agreement with the data.
In Figure~\ref{fig18}, $\big(\frac{\chi^2}{d.o.f}\big)_{\texttt{HERWIRI}}=2.03$ and $\big(\frac{\chi^2}{d.o.f}\big)_{\texttt{HERWIG}}=2.76$ For $\Delta R({j_{1},j_{2}})<4.2$.
\subsubsection{Dijet Angular Separation Distribution}
\begin{figure}[H]
\centering
\includegraphics[scale=0.4]{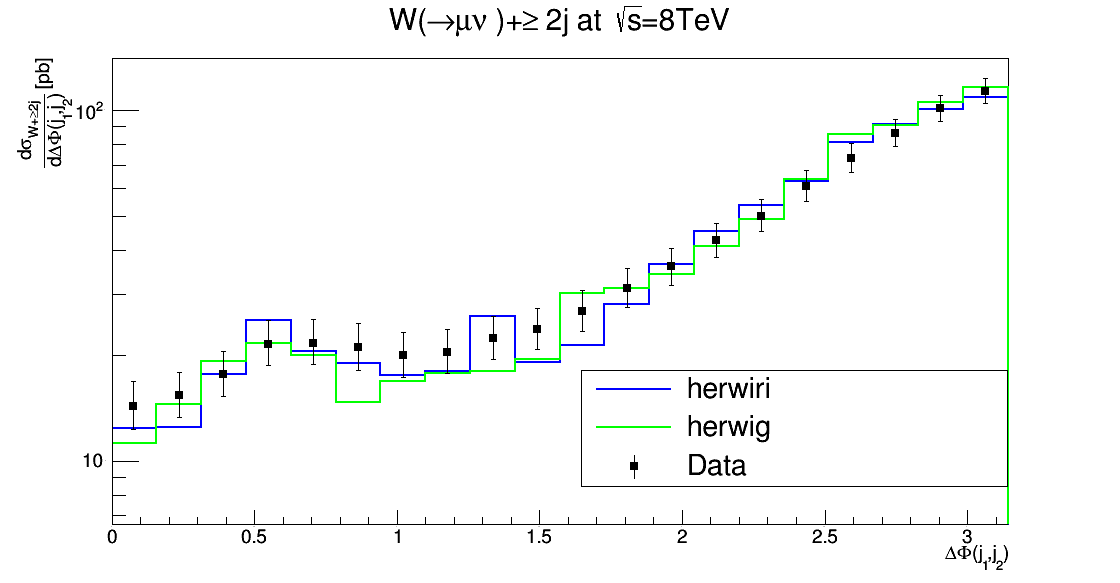}
\caption{Cross section for the production of W~+ jets as a function of the difference in azimuthal angle between the two leading jets $\Delta\Phi(j_{1},j_{2})$ in $N_{jet}\geq 2.$ The data are compared to predictions from MADGRAPH5\_aMC@NLO/HERWIRI1.031 and MADGRAPH5\_aMC@NLO/HERWIG6.521.}
\label{fig19}
\end{figure}

The differential cross section is given as a function of the difference in azimuthal angle $\Delta\Phi(j_{1},j_{2})$ for an inclusive jet multiplicity 2. In Figure~\ref{fig19}, the data are well modeled by the predictions provided by HERWIRI. In Figure~\ref{fig19}, $\big(\frac{\chi^2}{d.o.f}\big)_{\texttt{HERWIRI}}=0.81$ and $\big(\frac{\chi^2}{d.o.f}\big)_{\texttt{HERWIG}}=0.97$.\\
\par
The azimuthal angular distribution between the first and second leading jet is defined as
\begin{equation}
    \cos(\Delta\Phi(j_{1},j_{2}))=\frac{P_{x}(j_{1})P_{x}(j_{2})+P_{y}(j_{1})P_{y}(j_{2})}{\sqrt{P^2_{x}(j_{1})+P^2_{y}(j_{1})}\sqrt{P^2_{x}(j_{2})+P^2_{y}(j_{2})}}
\end{equation}
\subsubsection{The Azimuthal Angular Distribution Between the Muon and The Leading Jet}
The differential cross sections are shown as functions of the azimuthal angle between the muon and the first three leading jets for inclusive jet multiplicities 1--3. The azimuthal angle between the muon and the leading jet is defined as
\begin{equation}
\cos(\Delta\Phi(\mu,j_{1}))=\frac{P_{x}(\mu)P_{x}(j_{1})+P_{y}(\mu)P_{y}(j_{1})}{\sqrt{P^2_{x}(\mu)+P^2_{y}(\mu)}\sqrt{P^2_{x}(j_{1})+P^2_{y}(j_{1})}},
\end{equation}
with
\begin{equation}
\left\{ \begin{array}{ll}
         \mu^{\mu}=(E_{\mu},P_{x}(\mu),P_{y}(\mu),P_{L}(\mu)),\\
       j^{\mu}_{1}=(E_{j_{1}},P_{x}(j_{1}),P_{y}(j_{1}),P_{L}(j_{1})),\\\end{array} \right.
\end{equation}
\begin{figure}[H]
\centering
\includegraphics[scale=0.4]{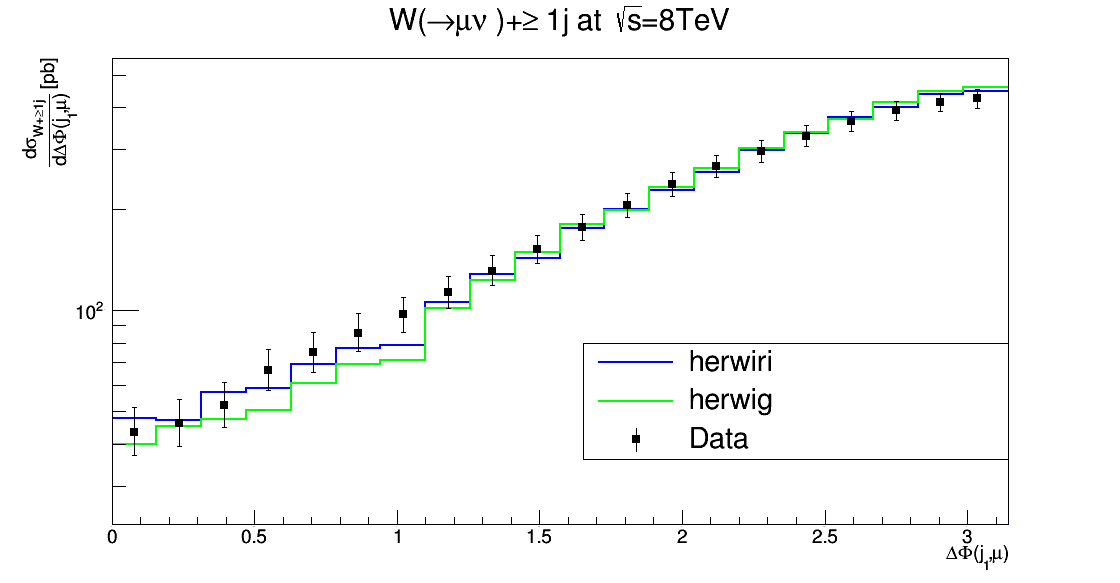}
\caption{Cross section for the production of W~+ jets as a function of the azimuthal angle between the muon and the leading jet $\Delta\Phi(\mu,j_{1})$ in $N_{jet}\geq 1.$ The data are compared to predictions from MADGRAPH5\_aMC@NLO/HERWIRI1.031 and MADGRAPH5\_aMC@NLO/HERWIG6.521.}
\label{fig20}
\end{figure}
\begin{figure}[H]
\centering
\includegraphics[scale=0.4]{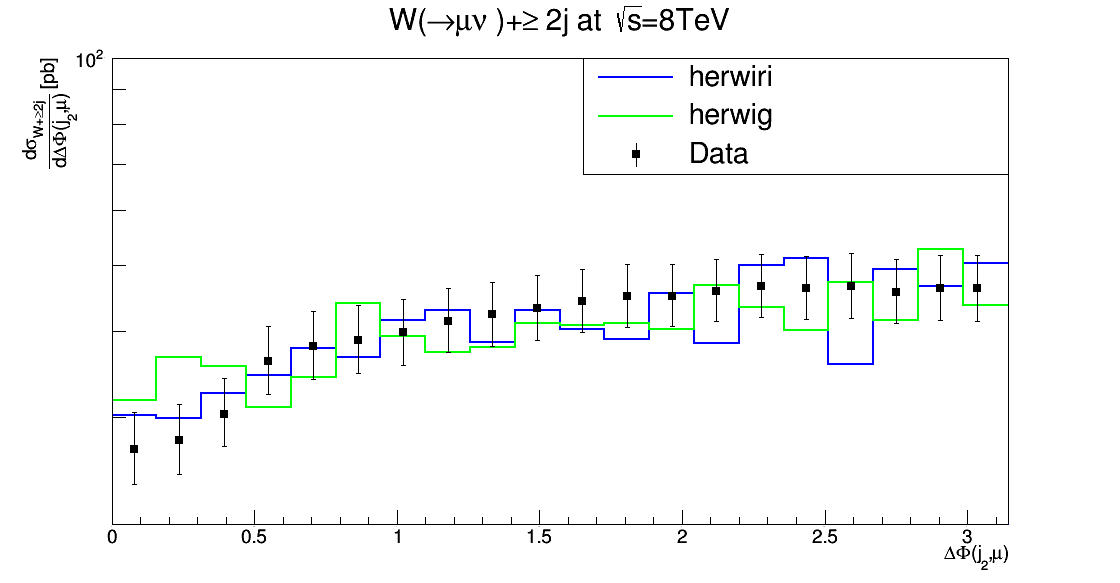}
\caption{Cross section for the production of W~+ jets as a function of  the azimuthal angle between the muon and the second leading jet $\Delta\Phi(\mu,j_{2})$ in $N_{jet}\geq2.$ The data are compared to predictions from MADGRAPH5\_aMC@NLO/HERWIRI1.031 and MADGRAPH5\_aMC@NLO/HERWIG6.521.}
\label{fig21}
\end{figure}
In Figure~\ref{fig20}, Figure~\ref{fig21}, and Figure~\ref{fig22} the data are better modeled by the predictions provided by HERWIRI as expected. In Figure~\ref{fig20}, $\big(\frac{\chi^2}{d.o.f}\big)_{\texttt{HERWIRI}}=0.42$ and $\big(\frac{\chi^2}{d.o.f}\big)_{\texttt{HERWIG}}=0.98$. In Figure~\ref{fig21}, $\big(\frac{\chi^2}{d.o.f}\big)_{\texttt{HERWIRI}}=0.80$ and $\big(\frac{\chi^2}{d.o.f}\big)_{\texttt{HERWIG}}=1.30$. In Figure~\ref{fig22}, $\big(\frac{\chi^2}{d.o.f}\big)_{\texttt{HERWIRI}}=0.92$ and $\big(\frac{\chi^2}{d.o.f}\big)_{\texttt{HERWIG}}=0.95$.\par

\begin{figure}[H]
\centering
\includegraphics[scale=0.4]{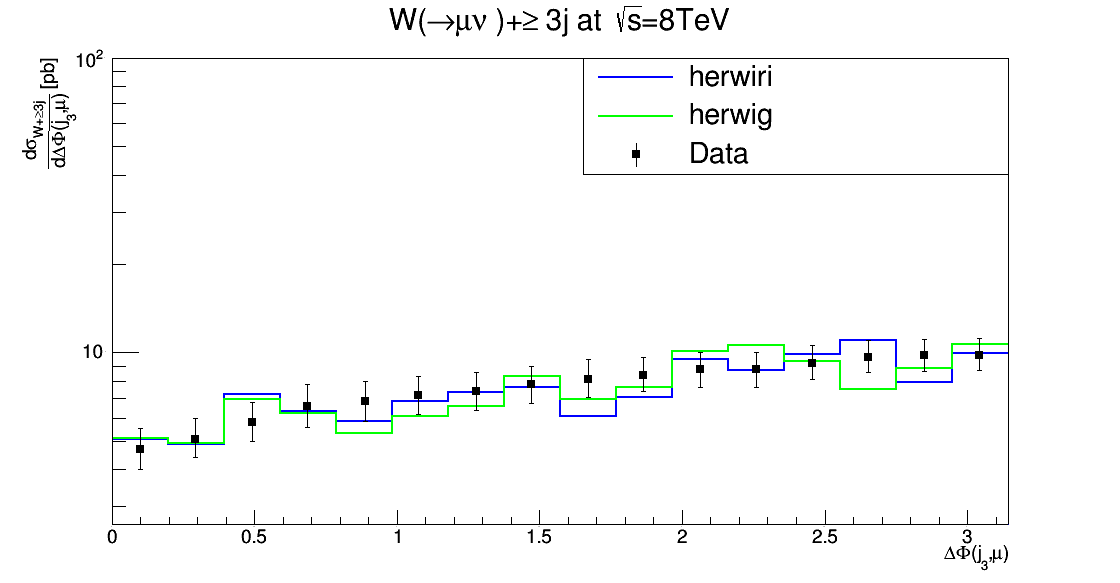}
\caption{Cross section for the production of W~+ jets as a function of the azimuthal angle between the muon and the second leading jet $\Delta\Phi(\mu,j_{3})$ in $N_{jet}\geq 3.$ The data are compared to predictions from MADGRAPH5\_aMC@NLO/HERWIRI1.031 and MADGRAPH5\_aMC@NLO/HERWIG6.521.}
\label{fig22}
\end{figure}

\subsubsection{Cross Sections}
\begin{figure}
\centering
\includegraphics[scale=0.4]{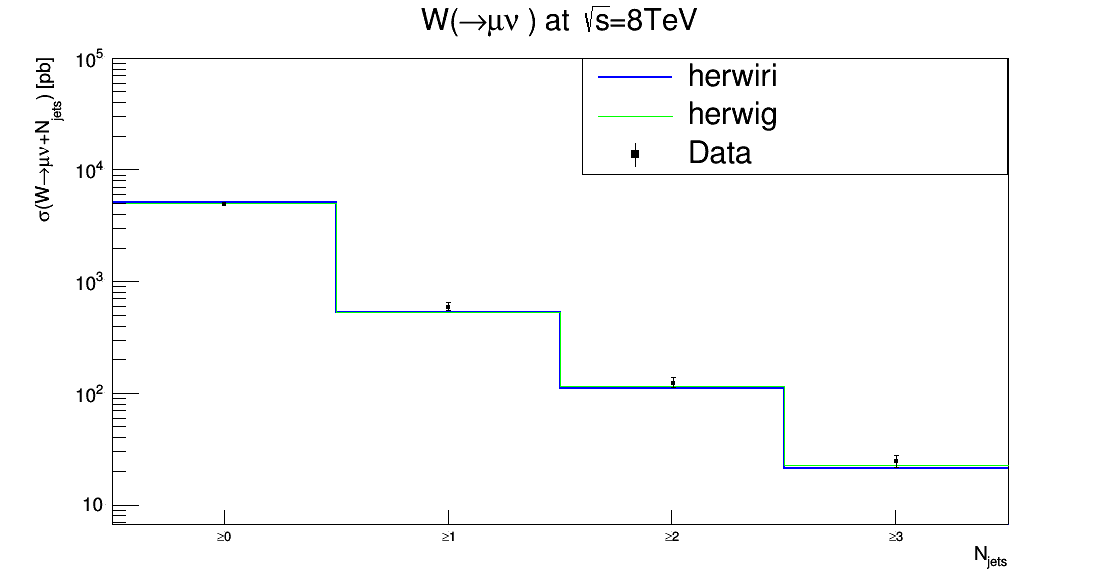}
\caption{Measured cross section versus inclusive jet multiplicity. The data are compared to predictions from MADGRAPH5\_aMC@NLO/HERWIRI1.031 and MADGRAPH5\_aMC@NLO/HERWIG6.521.}
\label{fig23}
\end{figure}

\begin{figure}[H]
\centering
\includegraphics[scale=0.4]{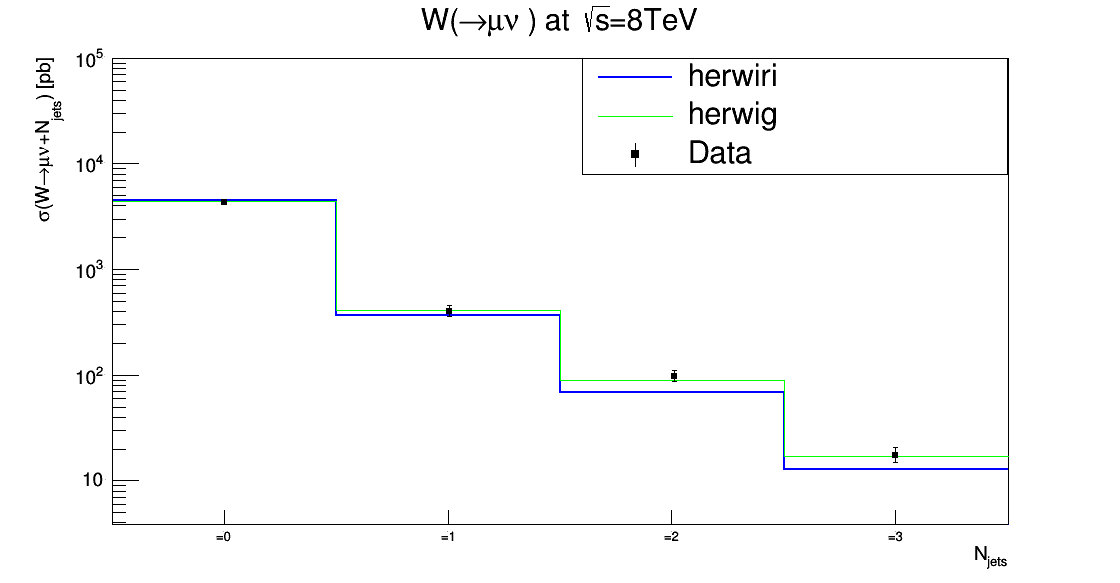}
\caption{Measured cross section versus exclusive jet multiplicity. The data are compared to predictions from MADGRAPH5\_aMC@NLO/HERWIRI1.031 and MADGRAPH5\_aMC@NLO/HERWIG6.521.}
\label{fig24}
\end{figure}

The measured $W(\rightarrow \mu\nu_{\mu})$~+~jets fiducial cross sections for inclusive and exclusive jet multiplicity distributions are shown in Figure~\ref{fig23} and Figure~\ref{fig24}, respectively. For inclusive jet multiplicity a good fit is given to the data by the theoretical predictions provided by HERWIRI and HERWIG. On the other hand, in Figure~\ref{fig24}, HERWIG gives a better fit to the measured cross sections for exclusive jet multiplicity 0--3. In Figure~\ref{fig23}, $\big(\frac{\chi^2}{d.o.f}\big)_{\texttt{HERWIRI}}=1.53$ and $\big(\frac{\chi^2}{d.o.f}\big)_{\texttt{HERWIG}}=1.49$. In Figure~\ref{fig24}, $\big(\frac{\chi^2}{d.o.f}\big)_{\texttt{HERWIRI}}=3.12$ and $\big(\frac{\chi^2}{d.o.f}\big)_{\texttt{HERWIG}}=1.09$.
\subsection{Results for $\sqrt{s}=$13 TeV}
\subsubsection{Transverse Momentum Distributions $P_{T}$ }

The differential cross sections in jet $P_{T}$ for inclusive jet multiplicities from 1 to 3 are shown and compared with predictions provided by HERWIRI and HERWIG.
\par
\begin{figure}[H]
\centering
\includegraphics[scale=0.4]{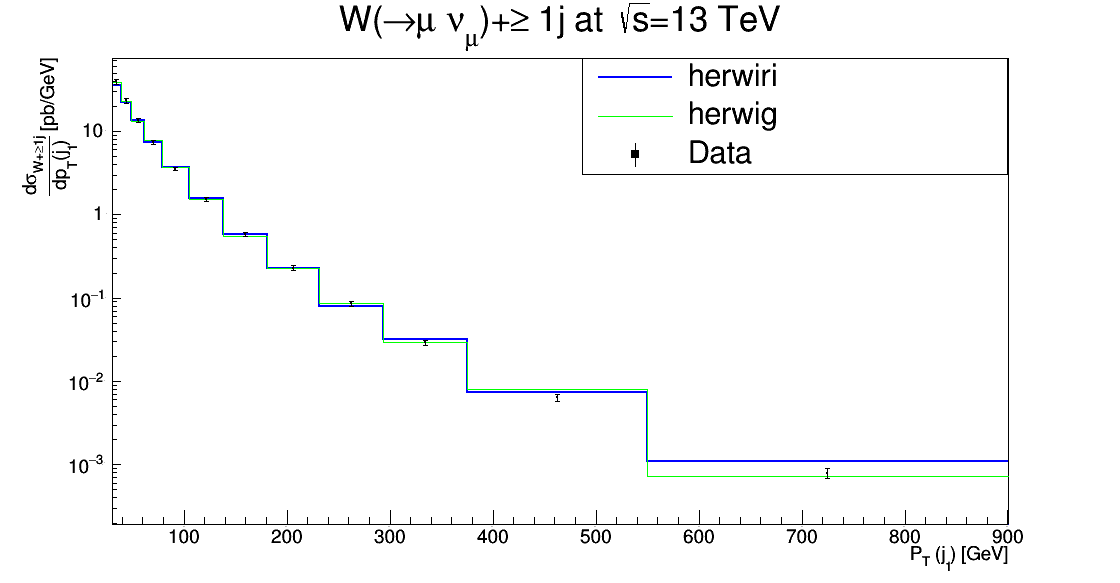}
\caption{Cross section for the production of W~+ jets as a function of the leading-jet $P_{T}$ in $N_{jet}\geq 1.$ The data are compared to predictions from MADGRAPH5\_aMC@NLO/HERWIRI1.031 and MADGRAPH5\_aMC@NLO/HERWIG6.521.}
\label{fig1a}
\end{figure}
\begin{figure}[H]
\centering
\includegraphics[scale=0.4]{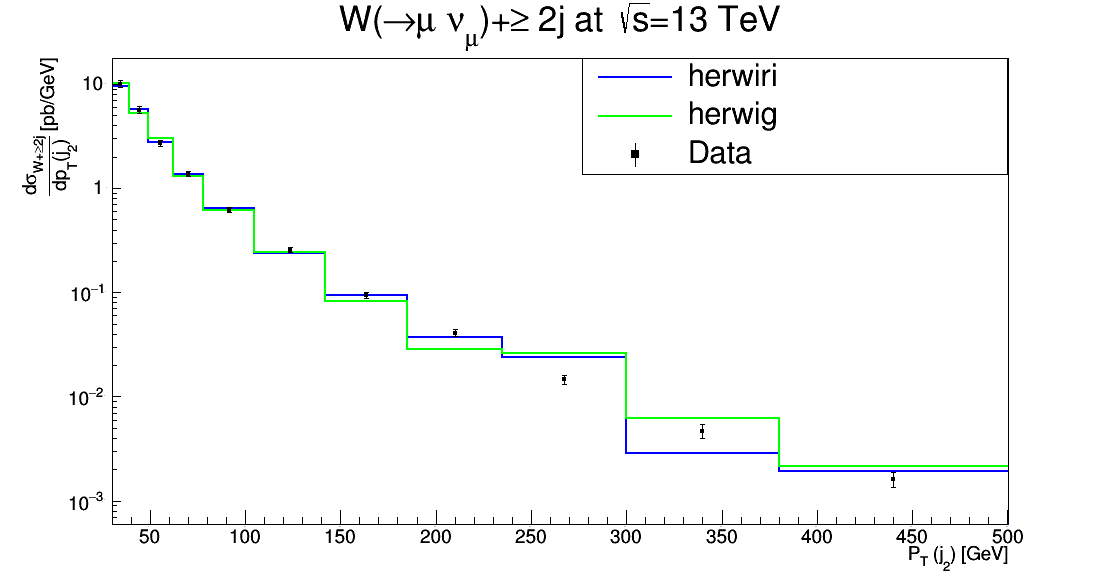}
\caption{Cross section for the production of W~+ jets as a function of the second leading-jet $P_{T}$ in $N_{jet}\geq2.$ The data are compared to predictions from MADGRAPH5\_aMC@NLO/HERWIRI1.031 and MADGRAPH5\_aMC@NLO/HERWIG6.521.}
\label{fig2a}
\end{figure}
\begin{figure}[H]
\centering
\includegraphics[scale=0.4]{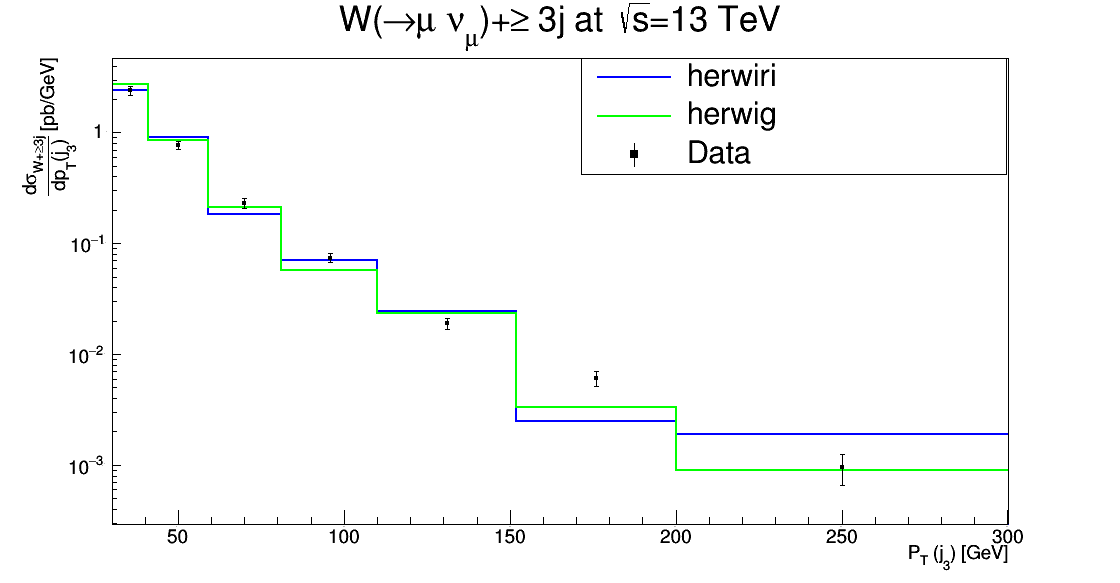}
\caption{Cross section for the production of W~+ jets as a function of the third leading-jet $P_{T}$ in $N_{jet}\geq 3.$ The data are compared to predictions from MADGRAPH5\_aMC@NLO/HERWIRI1.031 and MADGRAPH5\_aMC@NLO/HERWIG6.521.}
\label{fig3a}
\end{figure}

The differential cross sections as functions of the first three leading jets are shown in Figure~\ref{fig1a}, Figure~\ref{fig2a}, and Figure~\ref{fig3a}. In Figure~\ref{fig1a}, a good fit is provided by both HERWIRI and HERWIG for $P_{T}<181~\mathrm{GeV}$. In Figure~\ref{fig2a}, both HERWIRI and HERWIG provide a good fit to the data for $P_{T}<142~\mathrm{GeV}$. In Figure~\ref{fig1a}, for $P_{T}<181~\mathrm{GeV}$, $\big(\frac{\chi^2}{d.o.f}\big)_{\texttt{HERWIRI}}=0.77$ and $\big(\frac{\chi^2}{d.o.f}\big)_{\texttt{HERWIG}}=0.65$. In Figure~\ref{fig2a}  for $P_{T}<142$~GeV, $\big(\frac{\chi^2}{d.o.f}\big)_{\texttt{HERWIRI}}=0.98$ and $\big(\frac{\chi^2}{d.o.f}\big)_{\texttt{HERWIG}}=1.12$.\par

In Figure~\ref{fig3a}, a better fit is provided by HERWIRI to the data for $P_{T}<110~\mathrm{GeV}$ with $\big(\frac{\chi^2}{d.o.f}\big)_{\texttt{HERWIRI}}=1.21$ and $\big(\frac{\chi^2}{d.o.f}\big)_{\texttt{HERWIG}}=1.95$. In general, HERWIRI gives a better fit to the data. 
\subsubsection{The Absolute Rapidity Distribution $|Y(j)|$}
\begin{figure}[H]
    \centering
    \includegraphics[scale=0.4]{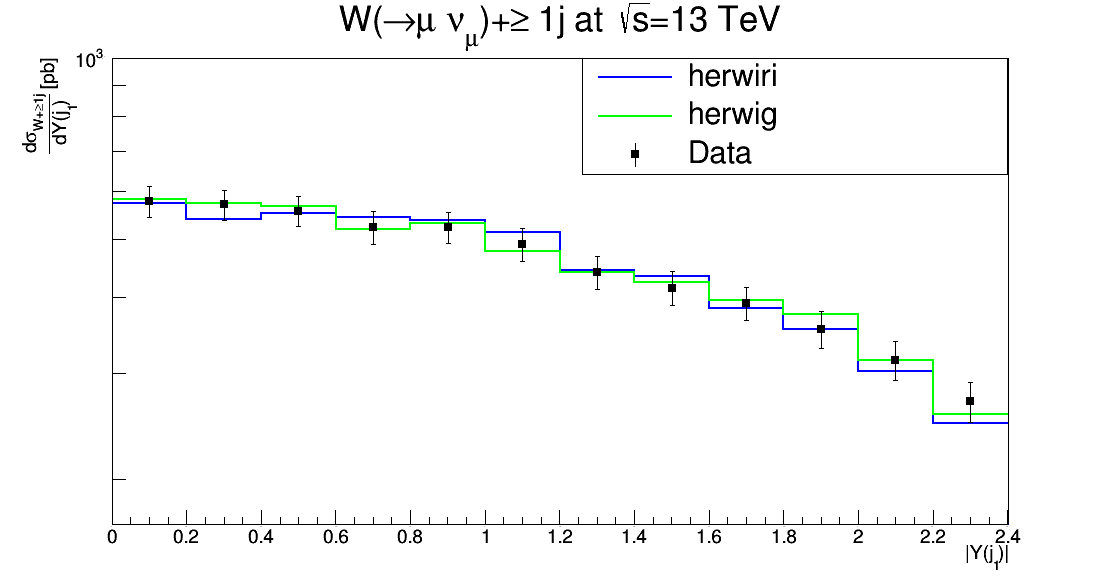}
    \caption{Cross section for the production of W~+ jets as a function of  $|Y(j_{1})|$ in $N_{jet}\geq 1.$ The data are compared to predictions from MADGRAPH5\_aMC@NLO/HERWIRI1.031 and MADGRAPH5\_aMC@NLO/HERWIG6.521.}
    \label{fig4a}
\end{figure}

\begin{figure}[H]
    \centering
    \includegraphics[scale=0.4]{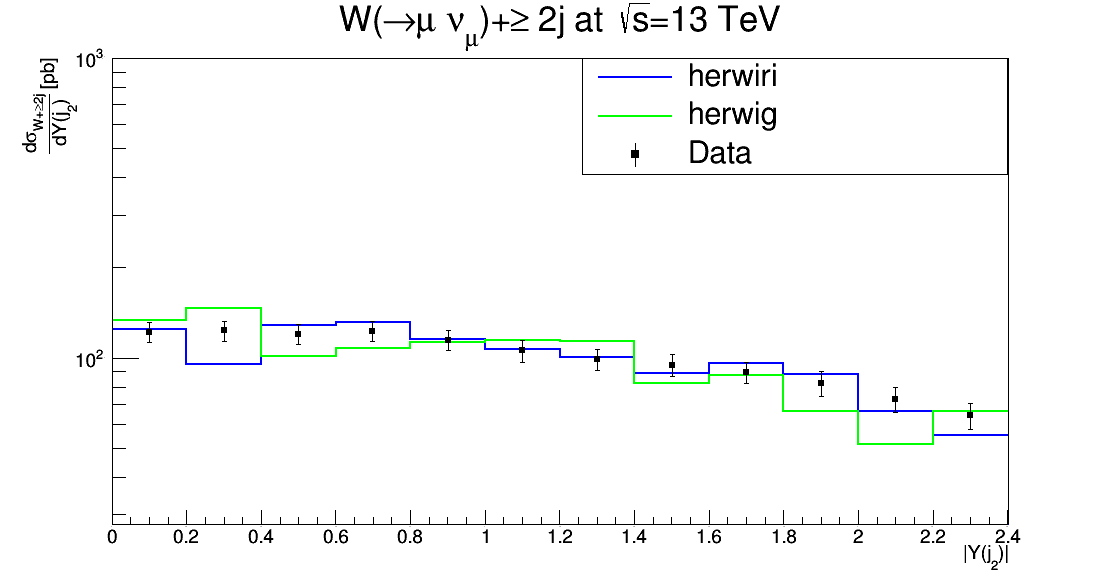}
    \caption{Cross section for the production of W~+ jets as a function of  $|Y(j_{2})|$ in $N_{jet}\geq 2.$ The data are compared to predictions from MADGRAPH5\_aMC@NLO/HERWIRI1.031 and MADGRAPH5\_aMC@NLO/HERWIG6.521.}
    \label{fig5a}
\end{figure}

\begin{figure}[H]
    \centering
    \includegraphics[scale=0.4]{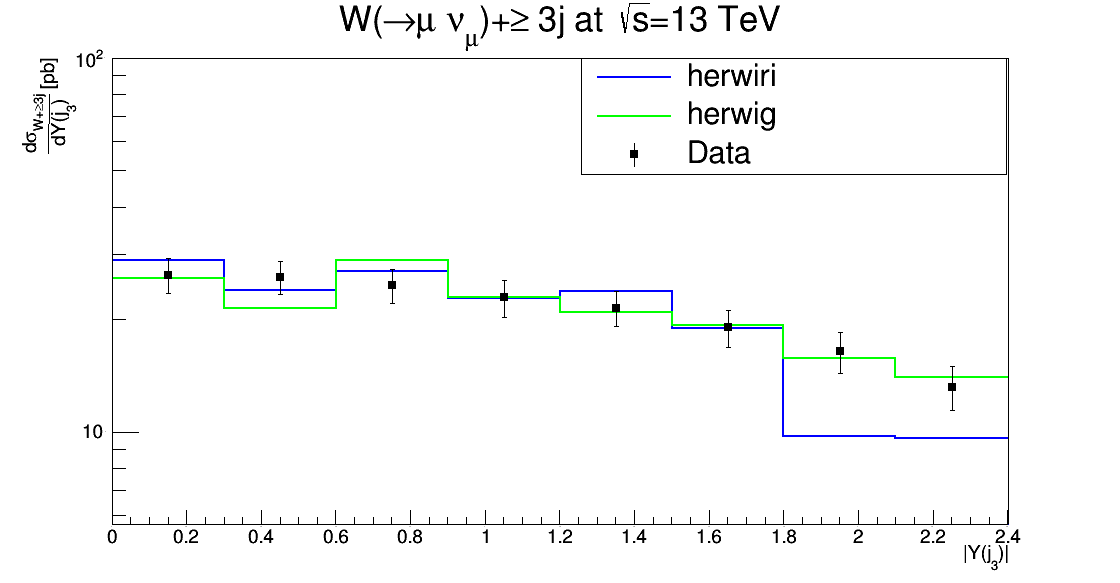}
    \caption{Cross section for the production of W~+ jets as a function of  $|Y(j_{3})|$ in $N_{jet}\geq 3.$ The data are compared to predictions from MADGRAPH5\_aMC@NLO/HERWIRI1.031 and MADGRAPH5\_aMC@NLO/HERWIG6.521.}
    \label{fig6a}
\end{figure}

The differential cross sections as functions of the absolute rapidities of the three leading jets are shown in Figure~\ref{fig4a}, Figure~\ref{fig5a}, and Figure~\ref{fig6a}. In Figure~\ref{fig4a}, a very good fit to the data is provided by HERWIRI and HERWIG with $\big(\frac{\chi^2}{d.o.f}\big)_{\texttt{HERWIRI}}=0.38$ and $\big(\frac{\chi^2}{d.o.f}\big)_{\texttt{HERWIG}}=0.14$. In Figure~\ref{fig5a}, predictions provided by HERWIRI give a better fit to the data as expected. We found: $\big(\frac{\chi^2}{d.o.f}\big)_{\texttt{HERWIRI}}=1.47$ and $\big(\frac{\chi^2}{d.o.f}\big)_{\texttt{HERWIG}}=3.10$. In Figure~\ref{fig6a}, HERWIG predictions are in better agreement with the data with $\big(\frac{\chi^2}{d.o.f}\big)_{\texttt{HERWIRI}}=2.72$ and $\big(\frac{\chi^2}{d.o.f}\big)_{\texttt{HERWIG}}=0.83$.\par

\subsubsection{The Scalar Sum of Jet Transverse Momenta $H_{T}$ }
In this subsection, the differential cross sections are shown as function of $H_{T}$ for inclusive jet multiplicities 1--3. 
\begin{figure}[H]
\centering
\includegraphics[scale=0.4]{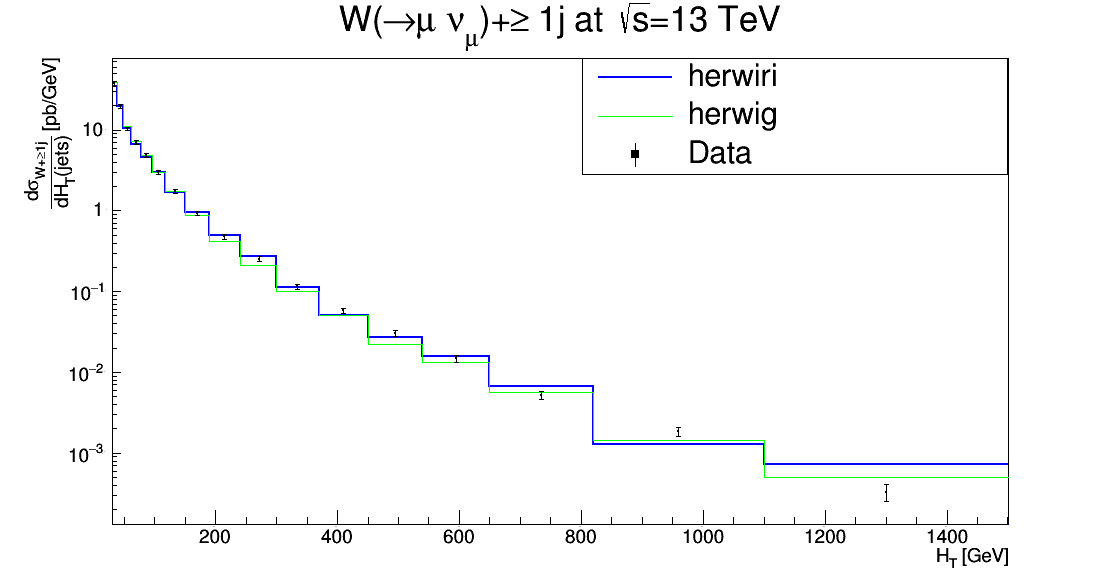}
\caption{Cross section for the production of W~+ jets as a function of  $H_{T}$ in $N_{jet}\geq 1.$ The data are compared to predictions from MADGRAPH5\_aMC@NLO/HERWIRI1.031 and MADGRAPH5\_aMC@NLO/HERWIG6.521.}
\label{fig7a}
\end{figure}

\begin{figure}[H]
\centering
\includegraphics[scale=0.4]{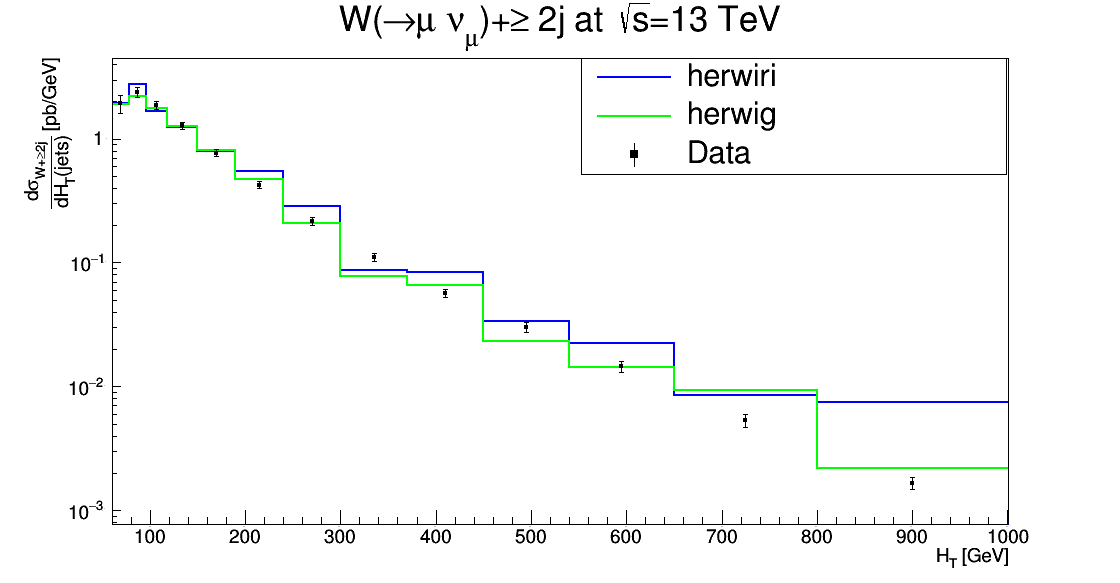}
\caption{Cross section for the production of W~+ jets as a function of  $H_{T}$ in $N_{jet}\geq2.$ The data are compared to predictions from MADGRAPH5\_aMC@NLO/HERWIRI1.031 and MADGRAPH5\_aMC@NLO/HERWIG6.521.}
\label{fig8a}
\end{figure}
\begin{figure}[H]
\centering
\includegraphics[scale=0.4]{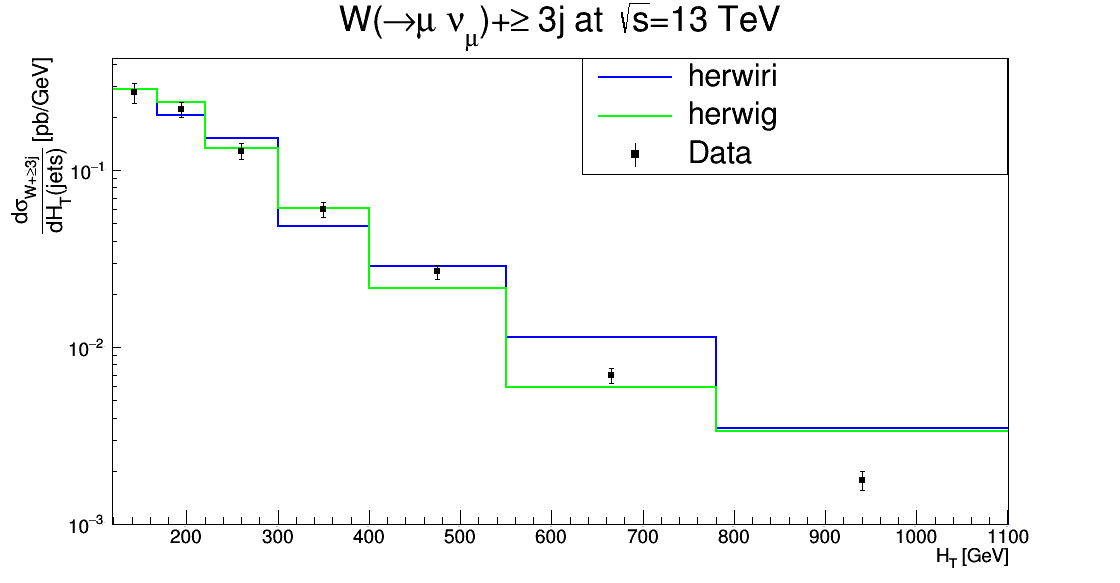}
\caption{Cross section for the production of W~+ jets as a function of $H_{T}$ in $N_{jet}\geq 3$. The data are compared to predictions from MADGRAPH5\_aMC@NLO/HERWIRI1.031 and MADGRAPH5\_aMC@NLO/HERWIG6.521.}
\label{fig9a}
\end{figure}
The differential cross sections as functions of $H_{T}$ for inclusive jet multiplicities 1--3 are shown in Figure~\ref{fig7a}, Figure~\ref{fig8a}, and Figure~\ref{fig9a}. In Figure~\ref{fig7a}, a very good fit is provided by HERWIRI and HERWIG predictions for $H_{T}<190~\mathrm{GeV}$. For higher values of $H_{T}$, HERWIG predictions are closer to the data. In Figure~\ref{fig8a}, in $H_{T}<190~\mathrm{GeV}$, HERWIG gives a better fit to the data. For $300<H_{T}~\mathrm{GeV}$, in some cases, HERWIRI predictions either overlap with the data or are closer to the data. In Figure~\ref{fig9a}, both HERWIRI and HERWIG predictions are in agreement with the data for $H_{T}<400~\mathrm{GeV}$.\\
In Figure~\ref{fig7a}, for $H_{T}<190~~\mathrm{GeV}$, $\big(\frac{\chi^2}{d.o.f}\big)_{\texttt{HERWIRI}}=0.26$ and $\big(\frac{\chi^2}{d.o.f}\big)_{\texttt{HERWIG}}=0.36$. In Figure~\ref{fig8a}, for $H_{T}<190~~\mathrm{GeV}$, $\big(\frac{\chi^2}{d.o.f}\big)_{\texttt{HERWIRI}}=1.21$ and $\big(\frac{\chi^2}{d.o.f}\big)_{\texttt{HERWIG}}=0.60$. In Figure~\ref{fig9a}, for $H_{T}<220~~\mathrm{GeV}$, $\big(\frac{\chi^2}{d.o.f}\big)_{\texttt{HERWIRI}}=0.61$ and $\big(\frac{\chi^2}{d.o.f}\big)_{\texttt{HERWIG}}=0.58$.\par

\subsubsection{The Azimuthal Angular Distribution Between the Muon and The Leading Jet}
The differential cross sections are shown as functions of the azimuthal angle between the muon and the first three leading jets for inclusive jet multiplicities 1--3. The azimuthal angle between the muon and the leading jet is defined as
\begin{equation}
\cos(\Delta\Phi(\mu,j_{1}))=\frac{P_{x}(\mu)P_{x}(j_{1})+P_{y}(\mu)P_{y}(j_{1})}{\sqrt{P^2_{x}(\mu)+P^2_{y}(\mu)}\sqrt{P^2_{x}(j_{1})+P^2_{y}(j_{1})}},
\end{equation}
with
\begin{equation}
\left\{ \begin{array}{ll}
         \mu^{\mu}=(E_{\mu},P_{x}(\mu),P_{y}(\mu),P_{L}(\mu)),\\
       j^{\mu}_{1}=(E_{j_{1}},P_{x}(j_{1}),P_{y}(j_{1}),P_{L}(j_{1})),\\\end{array} \right.
\end{equation}
\begin{figure}[H]
\centering
\includegraphics[scale=0.4]{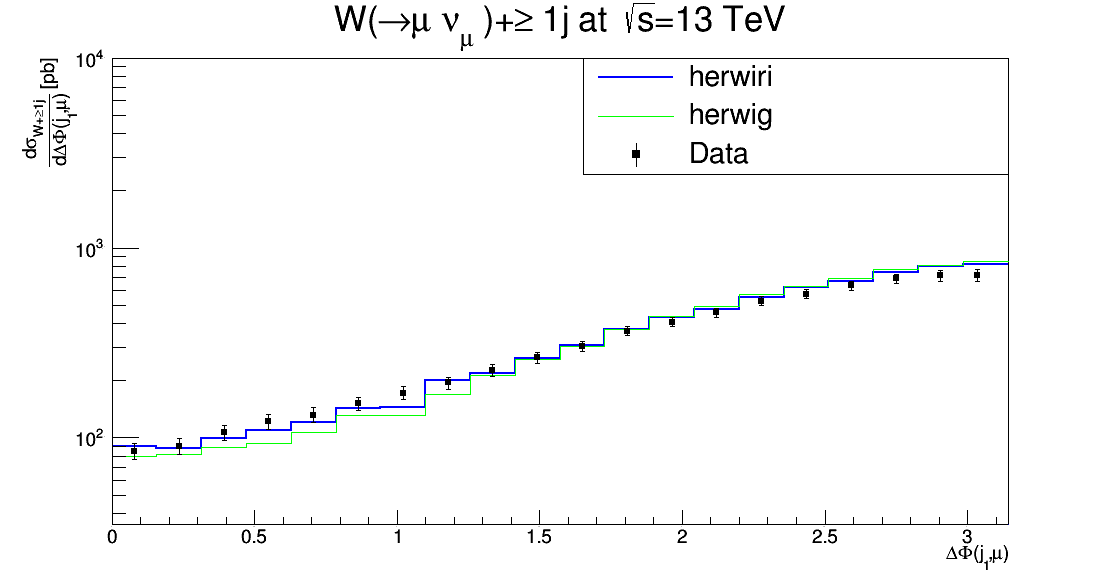}
\caption{Cross section for the production of W~+ jets as a function of the azimuthal angle between the muon and the leading jet $\Delta\Phi(\mu,j_{1})$ in $N_{jet}\geq 1.$ The data are compared to predictions from MADGRAPH5\_aMC@NLO/HERWIRI1.031 and MADGRAPH5\_aMC@NLO/HERWIG6.521.}
\label{fig10a}
\end{figure}
\begin{figure}[h!]
\centering
\includegraphics[scale=0.4]{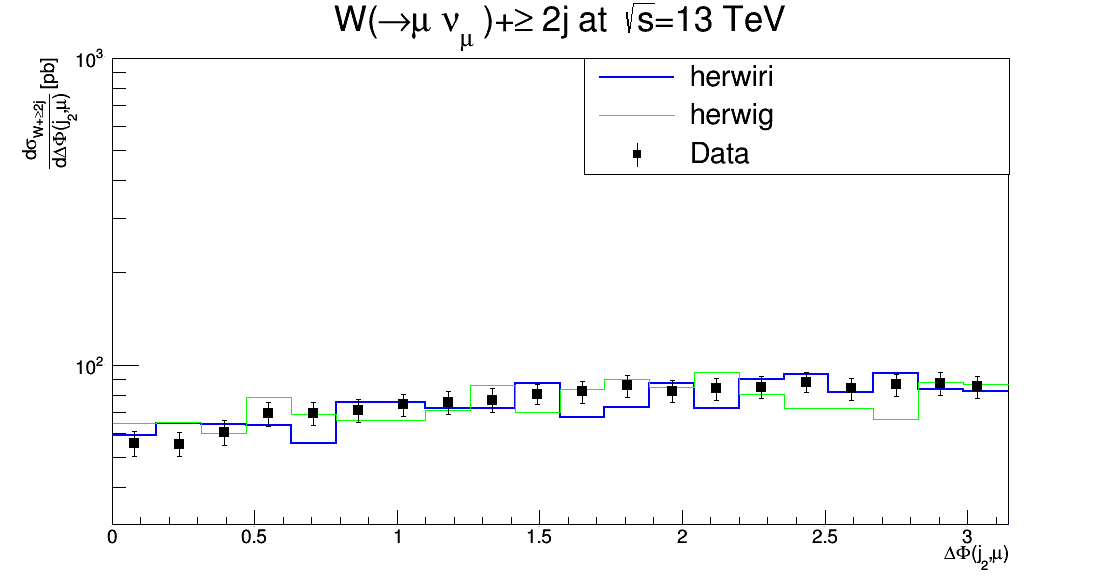}
\caption{Cross section for the production of W~+ jets as a function of  the azimuthal angle between the muon and the second leading jet $\Delta\Phi(\mu,j_{2})$ in $N_{jet}\geq2.$ The data are compared to predictions from MADGRAPH5\_aMC@NLO/HERWIRI1.031 and MADGRAPH5\_aMC@NLO/HERWIG6.521.}
\label{fig11a}
\end{figure}
\begin{figure}[H]
\centering
\includegraphics[scale=0.4]{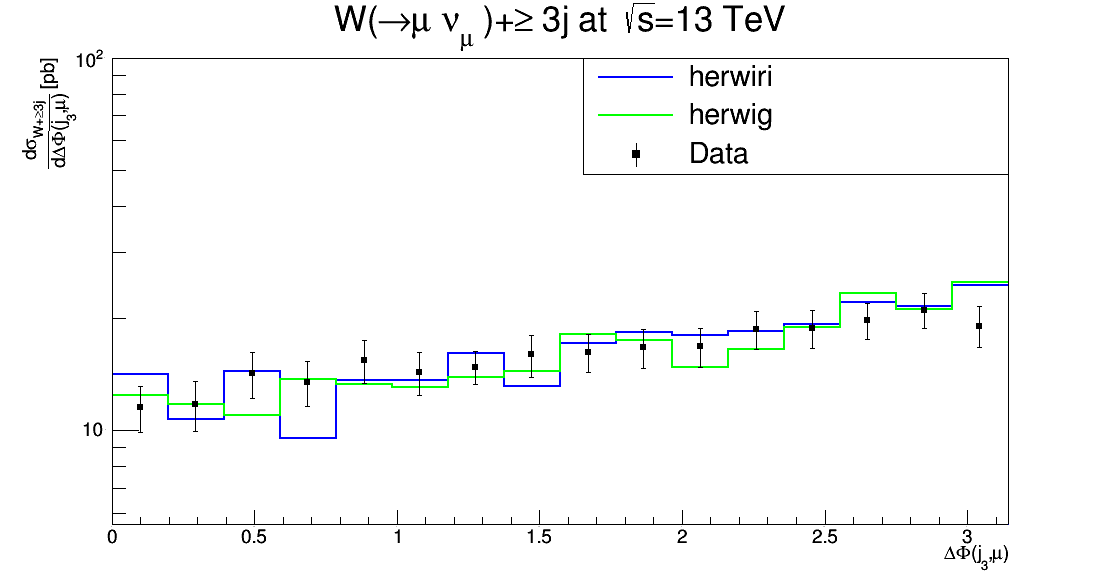}
\caption{Cross section for the production of W~+ jets as a function of the azimuthal angle between the muon and the second leading jet $\Delta\Phi(\mu,j_{3})$ in $N_{jet}\geq 3.$ The data are compared to predictions from MADGRAPH5\_aMC@NLO/HERWIRI1.031 and MADGRAPH5\_aMC@NLO/HERWIG6.521.}
\label{fig12a}
\end{figure}
\par
In Figure~\ref{fig10a} and Figure~\ref{fig11a}, the data are better modeled by the predictions provided by HERWIRI as expected. In Figure~\ref{fig10a}, $\big(\frac{\chi^2}{d.o.f}\big)_{\texttt{HERWIRI}}=1.14$ and $\big(\frac{\chi^2}{d.o.f}\big)_{\texttt{HERWIG}}=3.08$. In Figure~\ref{fig11a}, $\big(\frac{\chi^2}{d.o.f}\big)_{\texttt{HERWIRI}}=1.54$ and $\big(\frac{\chi^2}{d.o.f}\big)_{\texttt{HERWIG}}=1.73$.\\ 
\par
In Figure~\ref{fig12a}, predictions provided by HERWIRI and HERWIG are in fair agreement withe data. In three cases, HERWIG predictions overlap with the data while HERWIRI predictions either underestimate or overestimate the data. In two cases, HERWIRI predictions overlap with the data while HERWIG predictions in one case underestimates the data and in the other case overestimate the data. In general, in Figure~\ref{fig12a}, $\big(\frac{\chi^2}{d.o.f}\big)_{\texttt{HERWIRI}}=1.26$ and $\big(\frac{\chi^2}{d.o.f}\big)_{\texttt{HERWIG}}=1.12$.\par



\subsubsection{Cross Sections}
\begin{figure}[H]
\centering
\includegraphics[scale=0.4]{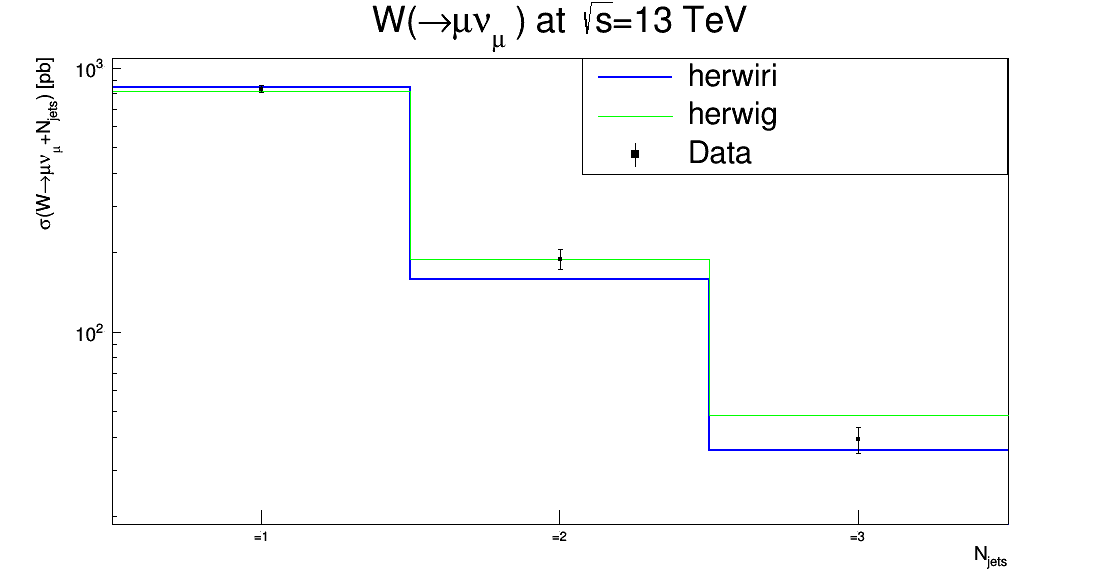}
\caption{Measured cross section versus exclusive jet multiplicity. The data are compared to predictions from MADGRAPH5\_aMC@NLO/HERWIRI1.031 and MADGRAPH5\_aMC@NLO/HERWIG6.521.}
\label{fig13a}
\end{figure}

\begin{figure}[H]
\centering
\includegraphics[scale=0.4]{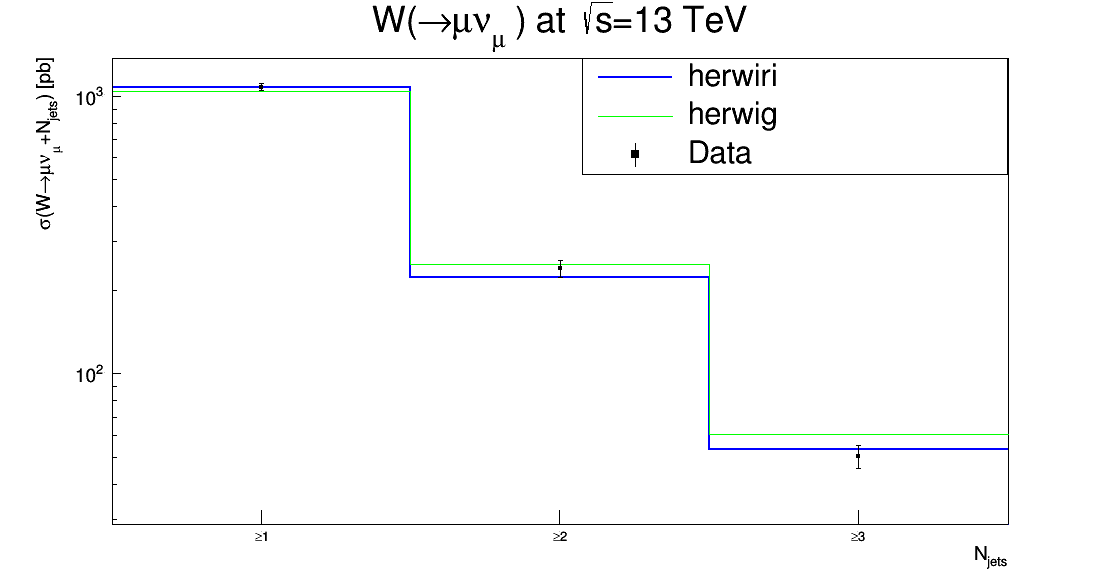}
\caption{Measured cross section versus inclusive jet multiplicity. The data are compared to predictions from MADGRAPH5\_aMC@NLO/HERWIRI1.031 and MADGRAPH5\_aMC@NLO/HERWIG6.521.}
\label{fig14a}
\end{figure}

The measured $W(\rightarrow \mu\nu_{\mu})$~+~jets fiducial cross sections for exclusive and inclusive jet multiplicity distributions are shown in Figure~\ref{fig13a} and Figure~\ref{fig14a}, respectively. For exclusive jet multiplicity, in Figure~\ref{fig13a}, $\big(\frac{\chi^2}{d.o.f}\big)_{\texttt{HERWIRI}}=2.37$ and $\big(\frac{\chi^2}{d.o.f}\big)_{\texttt{HERWIG}}=2.36$. For $N_{jet}=1$ and $N_{jet}=3$, HERWIRI gives a better fit to the data while for $N_{jet}=2$, HERWIG gives a better fit to the data and the prediction provided by HERWIRI underestimates the data. In Figure~\ref{fig14a}, HERWIRI gives a better fit to the measured cross sections for inclusive jet multiplicity 1--3 in general with: $\big(\frac{\chi^2}{d.o.f}\big)_{\texttt{HERWIRI}}=0.80$ and $\big(\frac{\chi^2}{d.o.f}\big)_{\texttt{HERWIG}}=2.89$. For $N_{jet}\geq1$ and $N_{jet}\geq3$, HERWIRI gives a better fit to the data while for $N_{jet}\geq2$, predictions provided by HERWIRI and HERWIG are in agreement with the data. 

\section{Study of the Energy Dependence of the results}
In this section we study the energy dependence of the results presented here and in I for similar distributions, e.g., $P_{T}$, $H_{T}$, $\eta$, etc. In each subsection, a comparison is performed between 7 TeV and 8 TeV, 7 TeV and 13 TeV, and 8 TeV and 13 TeV respectively. Needles to mention that for $P_{T}$ and $H_{T}$ distributions, the numerical value of $\frac{\chi^2}{d.o.f.}$ are calculated in the soft regime defined throughout this paper and I. 

\subsection{$P_{T}$ Distributions}
The 8 TeV $P_{T}$ results should be compared with the corresponding ones at 7 TeV in I, i.e., Figures 1, 4 and 6 for the ATLAS data, and Figures 27, 28, and 29 for the CMS data. As our calculations reveal, at 8 TeV the predictions provided by HERWIRI, compared to those provided by HERWIG, in all cases are in better agreement with the data within the associated soft regime. At 7 TeV, HERWIRI gives a  better fit than does HERWIG in all three ATLAS results for the soft regimes of Figs. 1, 4, and 6 of I. In the other (CMS) cases, while HERWIRI gives a comparable fit to that of HERWIG without the need of the intrinsic 2.2 GeV Gaussian $p_{T}$ distribution, the actual values of the respective $\frac{\chi^2}{d.o.f.}$ are lower for HERWIG. Unlike 7 TeV, in all cases at 8 TeV theoretical predictions provided by HERWIRI are in better agreement with the data.\par

Let us compare the 13 TeV results with our previous results in I at 7 TeV. In ATLAS data (Figures 1,4, and 6), for $W+\geq 1j$, $W+\geq 2j$, and $W+\geq 3j$ cases,  HERWIRI, compared to HERWIG, provides a better fit to the data. However, at 13 TeV, for $W+\geq nj$, $n=2,3$, HERWIRI's predictions are in better agreement with the data than those at 7 TeV with ATLAS data. In CMS at 7 TeV, i.e, figures 27, 28, and 29 in I, only for $W+\geq 1j$ HERWIRI gives a better fit to the data than it does for the corresponding ATLAS data but HERWIG gives an even better fit, with $\big(\frac{\chi^2}{d.o.f}\big)_{\texttt{HERWIRI}}=0.64$ and $\big(\frac{\chi^2}{d.o.f}\big)_{\texttt{HERWIG}}=0.35$. For both  $W+\geq 2j$, and $W+\geq 3j$ cases, HERWIG predictions are also in better agreement with the data.\par

Comparison between 8 TeV and 13 TeV reveals that for cases $W+\geq 1j$ and $W+\geq 2j$ at 13 TeV, HERWIRI gives a comparable or better fit to the data with lower numerical value of $\frac{\chi^2}{d.o.f.}$. For $W+\geq 3j$ HERWIRI still is in better agreement with data but with higher $\frac{\chi^2}{d.o.f.}$ numerical value. 

\subsection{$H_{T}$ Distributions}
We compare the 7, 8 and 13 TeV comparisons of theoretical and experimental results on $H_T$ as follows.
If we look at the $W+\geq 1j$ results, soft region $\frac{\chi^2}{d.o.f}$ for the  HERWIRI (HERWIG) fits are 0.28(1.94) for the ATLAS 7 TeV data (Fig. 20 in I), 0.57(0.40) for the CMS 7 TeV data (Fig. 30 in I), 0.56 (1.58) in the 8 TeV data in Fig. 4 here, and 0.26 (0.36) in the 13 TeV data in Fig. 31 here.   We see that in this case,  the HERWIRI fit is better for all three energies and that there is no evidence that it significantly degrades with increasing energy. 
    When we turn to the $W+\geq 2j$ case, we see that we have the correspnding soft region $\frac{\chi^2}{d.o.f}$ for HERWIRI
(HERWIG) as 2.96 (1.65) for the 7 TeV ATLAS data in Fig. 21 in I, 1.70 (1.36) for the 7 TeV CMS data in Fig. 31 in I, 0.53 (0.82) for the 8 TeV data in Fig. 5 here, and 1.21 (0.60) for the 13 TeV data in Fig. 32 here.   
We see that while HERWIRI still gives a comparable fit to HERWIG without the intrinsic 2.2 GeV Gaussion $p_T$ distribution, both fits are not as quite as good at 7 TeV as they are at 8 and 13 TeV. This is consistent with an increasing of the role of the soft radiation as we pass from $W+\geq 1j$ to $W+\geq 2j$. 
   The $W+\geq 3j$ case is consistent with our comment on the role of soft radiation, since in this case the corresponding results for the soft region $\frac{\chi^2}{d.o.f}$ fits for HERWIRI (HERWIG) are 3.80 (1.05) for the 7 TeV ATLAS data in Fig. 23 in I, 4.02 (4.32) for the 7 TeV CMS data in Fig. 32 in I, 13.2 (10.43) for the 8 TeV data in Fig. 6 here, and 0.61 (0.58) for the 13 TeV data in Fig. 33 here, respectively. These results show that the fits are generally better at the highest energy\footnote{At 8 TeV, the lowest soft regime bin given for the data does not really match either of the calculations. This is under study.}.

\subsection{$\eta(j)$ Distribution }
We then compare the $\eta(j)$ results presented for 8 TeV with the CMS data in I, i.e., Figures 33, 34, and 35. We see that at 7 TeV the $\frac{\chi^2}{d.o.f.}$ for the HERWIRI (HERWIG) fit to the $W+\geq 1j$ data in Fig. 33 in I is 0.39 (0.79), to the $W+\geq 2j$ data in Fig. 34 in I is 1.94 (1.71), and to the $W+\geq 3j$ data in Fig. 35 in I is 0.82 (0.61) whereas at 8 TeV the corresponding results are 0.30 (0.38) for the data in Fig. 7 here, 0.84 (0.66) for the data in Fig. 8 here, and 0.62 (1.02) for the data in Fig. 9 here. The increase in energy results in better fits in both cases in general. It also enhances a bit the relative quality of the the IR-improved fits compared to the unimproved ones.

\subsection{The Azimuthal Angular Distribution Between the Muon and The Leading Jet}

Comparison between the 8 TeV results and corresponding ones in I, i.e., Figures 36, 37, and 38, reveals that, unlike 8 TeV, only for $W+\geq 1j$ and $W+\geq 3j$ cases HERWIRI gives a good fit to the data. For $W+\geq 2j$, HERWIG results are in better agreement with the data, with $\big(\frac{\chi^2}{d.o.f}\big)_{\texttt{HERWIRI}}=2.73$ and $\big(\frac{\chi^2}{d.o.f}\big)_{\texttt{HERWIG}}=1.48$.\par

We then compare the 13 TeV results with CMS results at 7 TeV in I (see Figures 36, 37, and 38). At 7 TeV, only for $W+\geq 1j$, HERWIRI predictions give a better fit to the data with $\big(\frac{\chi^2}{d.o.f}\big)_{\texttt{HERWIRI}}=1.26$ and $\big(\frac{\chi^2}{d.o.f}\big)_{\texttt{HERWIG}}=2.67$. For $W+\geq 2j$, unlike 8 TeV, 7 TeV predictions provided by HERWIG give a better fit to the data, while at 13 TeV HERWIRI's predictions give a better fit to the data. For $W+\geq 3j$, unlike 8 TeV, both HERWIRI and HERWIG predictions give a good fit to the data at both 7 and 13 TeV. 

Finally, we compare the 13 TeV results with the ones at 8 TeV. In general, the results provided by HERWIRI are in better agreement with the data; however, the numerical values of $\frac{\chi^2}{d.o.f}$ are lower at 8 TeV. 

\subsection{Absolute Value of Y(j) distribution }
We thus compare the 13 TeV results with the corresponding ones in ATLAS (see Figures 7, 8, and 9 in I). At 7 TeV, only for $W+\geq 1j$, HERWIRI results give a better fit to the data than do those of HERWIG. At 7 TeV, both HERWIRI and HERWIG give good fits to the data. At 13 TeV, HERWIRI gives good fits to the data except for the $W+\geq 3j$ case where the $\frac{\chi^2}{d.o.f}$ is 2.72, while HERWIG gives a good fit to the data except for the $W+\geq 2j$ case where the $\frac{\chi^2}{d.o.f}$ is 3.10. 
    We conclude that there is a general consistency of the $\frac{\chi^2}{d.o.f}$ for the $W+\geq 1j$ case but that, for the other two cases, the $\frac{\chi^2}{d.o.f}$ are generally larger at 13 TeV.

\section{Summary}
In this paper the differential cross sections for different observables are presented at $\sqrt{s}=$ 8 and 13 TeV, and compared with the data. We also compared the results presented here with the results presented in I in terms of their energy dependence. As discussed in Section 4, at higher energy scale, results provided with HERWIRI in many cases are in better agreement with the data or they provide a lower value of $\frac{\chi^2}{d.o.f}$. At $\sqrt{s}=$8 TeV, in 19 cases out of 22, HERWIRI results either give a better fit to the data or are comparable with results provided with HERWIG. The same argument can be made at $\sqrt{s}=$13 TeV in which for 10 cases out of 12, HERWIRI results give a better fit to the data. \par
As we saw at 7 TeV, at 8 and 13 TeV the realization of the IR-improved DGLAP-CS theory, when used in the MADGRAPH5\_aMC@NLO/HERWIRI1.031 $\mathcal{O}(\alpha_{s})$ ME-matched parton shower framework, provides us with the opportunity to obtain a comparable or better fit to the data relative to the fit with the unimproved shower, in the soft regime, for the differential cross sections for a W boson produced in association with jets in pp collisions in the recent LHC results from CMS, without the need of an ad hoc hard intrinsic Gaussian distribution with an rms value of PTRMS = 2.2 GeV in the parton’s wave function. The results presented in this paper, taken together with those in I, clearly demonstrate a conceptual basis for the phenomenological correctness of such a Gaussian kick while using the usual IR-unimproved DGLAP-CS showers, i.e. HERWIG6.5.\

Specifically, we see that the results at 8 and 13 TeV show agreement with the findings in I at 7 TeV. We conclude that the effects of IR-improvement in jet observables in the production of W+ jets at the LHC are robust to changes in the cms energy. This raises similar expectations for the data at the FCC-hh, should it ever come to fruition. Such expectations will be discussed further elsewhere~\cite{elswh}.
\clearpage
\appendix

\section{Scale Factors for Theoretical Predictions}
\begin{table}[h!]
\centering
\begin{tabular}{ccccc} 
 \hline
 Figure number & $\alpha_{\texttt{HERWIRI}}$ & $\alpha_{\texttt{HERWIG}}$&{\Large\strut}$\big(\frac{\chi^2}{d.o.f}\big)_{\texttt{HERWIRI}}$&$\big(\frac{\chi^2}{d.o.f}\big)_{\texttt{HERWIG}}$ \\ [0.8ex] 
 \hline
 Figure 1 & ~~0.0523~~ & ~~0.0537~~&~~0.59~~&~~1.18~~ \\

 Figure 2 & 0.069 & 0.074 &~~1.02~~&~~1.95~~ \\
 
 Figure 3 & 0.03212 & 0.03111  & ~~1.47~~&~~2.04~~ \\

 Figure 4& 0.05671 & 0.05721 &~~0.56~~&~~1.48~~ \\

 Figure 5& 0.0701 & 0.0751 &~~0.53~~&~~0.82~~ \\ 

 Figure 6& 0.0351 & 0.03092  &~~13.20~~&~~10.43~~ \\

 Figure 7&  0.052016&  0.052775 &~~0.30~~&~~0.38~~\\

 Figure 8& 0.06923 & 0.07114&~~0.84~~&~~0.66~~\\
 
 Figure 9& 0.02769 & 0.028416 &~~0.62~~&~~1.02~~\\

 Figure 10&0.03445 & 0.03287  &~~1.17~~&~~1.43~~\\

 Figure 11& 0.0670 & 0.0719 &~~1.67~~&~~1.92~~\\

 Figure 12& 0.07566 & 0.07701 &~~2.00~~&~~1.98~~\\

 Figure 13& 0.03193 &0.03173& 0.48 & 1.04  \\

 Figure 14& 0.031543 & 0.033212 &1.20& 0.56 \\

 Figure 15& 0.033432 &  0.034312 &0.33 & 0.52 \\

 Figure 16& 0.0768 & 0.0786 & 0.92 & 1.01\\
 
 Figure 17&0.0310  & 0.0329 & 1.1 & 2.45\\

 Figure 18& 0.02432 & 0.02427 & 2.03 & 2.73\\
 
 Figure 19& 0.025132  &0.025212 & 0.81 & 0.97\\

 Figure 20& 0.0189112 &0.01888931 & 0.42 & 0.98 \\
 
 Figure 21& 0.0249110 &0.0245101 & 0.80 & 1.30\\

 Figure 22& 0.0103311 &0.0104312 & 0.92 & 0.95 \\

 \hline
\end{tabular}
\caption{Summary of the scale factors applied to the theoretical predictions for CMS at $\sqrt{s}=8$~TeV}
\label{Table3}
\end{table}
\newpage

\section{Scale Factors for Theoretical Predictions}
\begin{table}[h!]
\centering
\begin{tabular}{ccccc} 
 \hline
 Figure number & $\alpha_{\texttt{HERWIRI}}$ & $\alpha_{\texttt{HERWIG}}$&{\Large\strut}$\big(\frac{\chi^2}{d.o.f}\big)_{\texttt{HERWIRI}}$&$\big(\frac{\chi^2}{d.o.f}\big)_{\texttt{HERWIG}}$ \\ [0.8ex] 
 \hline
 Figure 25 & ~~0.202~~ & ~~0.191~~ &~~0.76~~&~~0.65~~\\

 Figure 26 & 0.5782 & 0.455 &~~0.98~~&~~1.02~~ \\
 
 Figure 27& 0.4916 & 0.368  & ~~1.21~~&~~1.95~~\\

 Figure 28& 0.210 & 0.195 &~~0.37~~&~~0.14~~  \\

 Figure 29& 0.5938 & 0.459 &~~1.47~~&~~3.11~~ \\ 

 Figure 30& 0.473 & 0.346  &~~2.725~~&~~0.83\\

 Figure 31&  0.203&  0.189 &~~0.26~~&~~0.36~~\\

 Figure 32& 0.615 & 0.46&~~1.21~~&~~0.58 \\
 
 Figure 33& 0.51 & 0.351 &~~0.61~~&~~0.58 \\

 Figure 34&0.209 & 0.185  &~~1.14~~&~~3.08  \\

 Figure 35& 0.5813 & 0.436 &~~1.54~~&~~1.73 \\

 Figure 36& 0.5014 & 0.371 &~~1.27~~&~~1.13  \\

 Figure 37& 0.6921 &0.565& 2.37 & 2.39  \\

 Figure 38& 0.632 & 0.558 &0.805& 2.89 \\

 \hline
\end{tabular}
\caption{Summary of the scale factors applied to the theoretical predictions for CMS at $\sqrt{s}=13$~TeV}
\label{Table4}
\end{table}


\clearpage

\begin{thebibliography}{99}

  \bibitem{bsh1} 
   B.~Shakerin and B.~F.~L.~Ward,
  ``IR-improved DGLAP parton shower effects in W + jets in pp collisions at $\sqrt{s}=7$ TeV,''
  Phys.\ Rev.\ D {\bf 100}, 034026 (2019)
    \bibitem{Joseph:2010cq} 
  S.~Joseph, S.~Majhi, B.~F.~L.~Ward and S.~A.~Yost,
  ``New Approach to Parton Shower MC's for Precision QCD Theory: HERWIRI 1.0(31),''
  Phys.\ Rev.\ D {\bf 81}, 076008 (2010)
       \bibitem{Alwall:2014hca} 
  J.~Alwall {\it et al.},
  ``The automated computation of tree-level and next-to-leading order differential cross sections, and their matching to parton shower simulations,''
  JHEP {\bf 1407}, 079 (2014)
  
      \bibitem{Corcella:2000bw} 
  G.~Corcella, I.~G.~Knowles, G.~Marchesini, S.~Moretti, K.~Odagiri, P.~Richardson, M.~H.~Seymour and B.~R.~Webber,
  ``HERWIG 6: An Event generator for hadron emission reactions with interfering gluons (including supersymmetric processes),''
  JHEP {\bf 0101}, 010 (2001)
  
    \bibitem{Khachatryan:2016fue} 
  V.~Khachatryan {\it et al.} [CMS Collaboration],
  ``Measurements of differential cross sections for associated production of a W boson and jets in proton-proton collisions at $\sqrt{s} =$ 8 TeV,''
  Phys.\ Rev.\ D
  [Phys.\ Rev.\ D {\bf 95}, 052002 (2017)]
  [arXiv:1610.04222 [hep-ex]].
  \\\texttt{https://www.hepdata.net/record/76995}
\bibitem{Sirunyan:2017wgx} 
  A.~M.~Sirunyan {\it et al.} [CMS Collaboration],
  ``Measurement of the differential cross sections for the associated production of a $W$ boson and jets in proton-proton collisions at $\sqrt{s}=13$ TeV,''
  Phys.\ Rev.\ D {\bf 96}, no. 7, 072005 (2017)
  [arXiv:1707.05979 [hep-ex]].
  \\\texttt{https://www.hepdata.net/record/ins1610623}
  
  \bibitem{Shakerin:2017hbd} 
  B.~Shakerin,
  ``IR-improved DGLAP-CS parton shower effects in W + jets at $\sqrt{s}$ = 7, 8, and 13 TeV.,`` Dissertation, Baylor University, 2017.
  
  

  

  

\bibitem{Cacciari:2011ma} 
  M.~Cacciari, G.~P.~Salam and G.~Soyez,
  ``FastJet User Manual,''
  Eur.\ Phys.\ J.\ C {\bf 72}, 1896 (2012)
  
  \bibitem{elswh} 
  B. Shakerin and B.F.L. Ward, unpublished.
  

\end{thebibliography}
\end{document}